\documentclass[11pt, a4paper]{article}

\usepackage{jcappub_ver}
\usepackage{enumerate}
\usepackage{amsbsy}
\usepackage{graphics}
\usepackage{mathrsfs}
\usepackage{wrapfig}

\newcommand{\be}{\begin{equation}} \newcommand{\ee}{\end{equation}}
\newcommand{\bea}{\begin{eqnarray}} \newcommand{\eea}{\end{eqnarray}}
\newcommand{\el}{\nonumber \\}
\newcommand{\re}[1]{(\ref{#1})}

\newcommand{\pat}{\partial}

\renewcommand{\sec}[1]{section \ref{#1}}
\newcommand{\fig}[1]{figure \ref{#1}}
\newcommand{\brt}[1]{[#1]}
\newcommand{\para}{\paragraph}

\newcommand{\LCDM}{$\Lambda$CDM\ }
\newcommand{\GN}{G_{\mathrm{N}}}

\newcommand{\rmd}{\mathrm{d}}

\newcommand{\nonum}{\\}
\newcommand{\etal} {et al.}

\newcommand{\adot}{\dot{a}}
\newcommand{\addot}{\ddot{a}}

\renewcommand{\H}{\frac{\adot}{a}}
\newcommand{\HH}{\frac{\adot^2}{a^2}}

\newcommand{\phit}{\tilde{\phi}}

\newcommand{\hfit}{h_{\mathrm{fit}}}
\newcommand{\wHtot}{w_{H\mathrm{tot}}}
\newcommand{\wDtot}{w_{D\mathrm{tot}}}
\newcommand{\rhom}{\rho_{\mathrm{m}}}

\newcommand{\av}[1]{\langle{#1}\rangle}
\newcommand{\sQ}{\mathcal{Q}}
\newcommand{\sR}{{^{(3)}R}}

\newcommand{\Om}{\Omega_{\mathrm{m}}}
\newcommand{\Omn}{\Omega_{\mathrm{m0}}}
\newcommand{\Omfit}{\Omega_{\mathrm{mfit}}}

\newcommand{\OQ}{\Omega_{\sQ}}
\newcommand{\OQn}{\Omega_{\sQ0}}
\newcommand{\OR}{\Omega_{R}}
\newcommand{\ORn}{\Omega_{R0}}
\newcommand{\OK}{\Omega_{K}}
\newcommand{\OKn}{\Omega_{K0}}

\newcommand{\sO}{\mathcal{O}}
\newcommand{\mB}{\mathscr{B}}
\newcommand{\mC}{\mathscr{C}}

\newcommand{\PRD}[1]{{\it Phys. Rev.} {\bf D#1}}

\newcommand{\PRL}[1]{{\it Phys. Rev. Lett.} {\bf #1}}

\newcommand{\PLA}[1]{{\it Phys. Lett.} {\bf A#1}}
\newcommand{\PLB}[1]{{\it Phys. Lett.} {\bf B#1}}
\newcommand{\MNRAS}[1]{{\it Mon. Not. Roy. Astron. Soc.} {\bf #1}}
\newcommand{\APJ}[1]{{\it Astrophys. J.} {\bf #1}}

\newcommand{\CQG}[1]{{\it Class. Quant. Grav.} {\bf #1}}
\newcommand{\GRG}[1]{{\it Gen. Rel. Grav.} {\bf #1}}
\newcommand{\AaA}[1]{{\it Astron. \& Astrophys.} {\bf #1}}
\newcommand{\PROG}[1]{{\it Prog. Theor. Phys.} {\bf #1}}

\newcommand{\IJMPD}[1]{{\it Int. J. Mod. Phys.} {\bf D#1}}

\title{Violation of the FRW consistency condition as a signature of backreaction}

\author[a]{C\'eline Boehm}

\author[b]{and Syksy R\"{a}s\"{a}nen}

\affiliation[a]{Institute for Particle Physics Phenomenology, Durham University \\
South Road, Durham, DH1 3LE, United Kingdom \\
and LAPTH, U. de Savoie, CNRS \\
BP 110, 74941 Annecy-Le-Vieux, France}

\affiliation[b]{University of Helsinki, Department of Physics \\
and Helsinki Institute of Physics \\
P.O. Box 64, FIN-00014 University of Helsinki, Finland}

\emailAdd{c {\it dot} m {\it dot} boehm {\it at} durham {\it dot} ac {\it dot} uk }

\emailAdd{syksy {\it dot} rasanen {\it at} iki {\it dot} fi}

\abstract{
We propose a backreaction toy model with realistic features
and confront it with the Union2.1 supernova data.
The model provides a good fit even though the
expansion history is quite different from the \LCDM model and the
effective equation of state is far from $-1$.
We discuss compatibility (or lack thereof) with
other observations. We show that the FRW consistency
condition between distance and expansion rate is
violated with an amplitude that is slightly below
the current observational limits.
We expect that this is also the case in the real
universe if backreaction is significant, providing
a distinct signature of backreaction.
}

\begin{document}
  
\maketitle

\setcounter{tocdepth}{2}

\setcounter{secnumdepth}{3}

\section{Introduction} \label{sec:intro}

\para{Observations.}

The observed average expansion rate and distance are at late times
about a factor of two higher than predicted by the homogeneous
and isotropic and spatially flat matter-dominated cosmological
model (see \cite{Buchert:2011} for discussion and references).
There are three possible explanations: an exotic matter component
with negative pressure, failure of general relativity on cosmological
scales or the breakdown of the validity of the homogeneous and
isotropic approximation.

Most observations are consistent with extending the
matter-dominated spatially flat Friedmann--Robertson--Walker (FRW)
model to the \LCDM model by simply adding vacuum energy. 
There are a few discrepancies, however. Temperature variations in
the cosmic microwave background (CMB) along lines of sight
that intersect density extrema of the large-scale structure
are much higher than expected \cite{ISWpower}, and there are also
other CMB anomalies on large angular scales \cite{asym}.
The distribution of galaxies on large scales appears
to be less homogeneous than expected, particularly when
it comes to extreme structures \cite{Einasto, inhom, 2dF},
though the overall amplitude of the two-point function is
consistent with the \LCDM model \cite{Contreras:2013}.

It has also been argued that observations of supernovae (SNe),
supplemented by other data, favour departures
from the \LCDM model in the form of more acceleration
today \cite{moreacc}, less acceleration (or even deceleration)
today \cite{lessacc} or more deceleration at early times
\cite{Perivolaropoulos:2008}.
The results depend strongly on the assumed parametrisation
of the cosmological model \cite{trans} and choice of light curve
fitter \cite{fitter}. None of these deviations are statistically
significant \cite{kinematic}, but the analyses show that
large departures from the \LCDM expansion history are allowed.
Currently, SNe are the most detailed probe of the
distance--redshift relation (from which the expansion rate is inferred),
but observations of large-scale structure, in particular 
the baryon acoustic oscillation (BAO) pattern
\cite{Bassett:2009, Gaztanaga:2008, Blake:2011, Reid:2012, Blake:2012, Chuang:2012, Xu:2012, BAOrad} (see also \cite{BAOquest}),
offer an increasingly accurate picture of distances and expansion
rates, though the analysis is more model-dependent.
Galaxy ages are a relatively new and important model-independent
way to measure the expansion rate, with systematics that
are not yet as well-studied as for more established probes
\cite{ages, Jimenez, Moresco:2012, Zhang:2012}.

\para{Backreaction.}

One possibility is that the factor two failure in the
predictions of the homogeneous and isotropic matter-dominated
model is related to the known breakdown of homogeneity
and isotropy due to structure formation, instead of new
fundamental physics
\cite{Buchert:2000, Wetterich:2001, Schwarz:2002, Rasanen, Kolb:2004}.
The average expansion rate is in general affected by
inhomogeneity and anisotropy, even if the
distribution of matter is statistically homogeneous and isotropic.
The effect of deviations from homogeneity and/or isotropy on
the average expansion rate is known as backreaction
\cite{Shirokov:1962, Buchert:1995, fitting, Buchert:1999a, Buchert:2001}
(see \cite{Ellis:2005, revs, Buchert:2011} for reviews).
Inhomogeneity can increase the average expansion rate, and
it can even lead to accelerating expansion in a dust universe
\cite{acc, LTBacc}.
In a statistically homogeneous and isotropic universe, faster
expansion rate implies larger redshifts and distances
\cite{Rasanen:2008b, Rasanen:2009b}.
The possibility that change in the average expansion rate
due to structure formation would explain the late-time observations
of faster expansion and longer distances
is called the backreaction conjecture.

Given the standard scenario of gravitational structure formation
from small Gaussian perturbations in baryons and cold
dark matter, there are no free parameters for backreaction.
The initial matter and radiation densities and the perturbation
spectrum is known model-independently from CMB observations
\cite{Vonlanthen:2010, Audren:2012}.
Therefore, the average expansion rate is fully determined in principle.
However, in practice it is difficult to calculate the effect
of non-linear structures.
In Newtonian cosmology \cite{Buchert:1995} and in
relativistic perturbation theory \cite{Rasanen:2011b}
(see also \cite{Green}), the effect is small, given certain assumptions.
In a semirealistic model, backreaction has been found to be
of the same order of magnitude as the observed change in the
expansion rate, given certain approximations \cite{peak}.
The magnitude of backreaction in the real universe is an open question.

However, backreaction can be tested by independent observations of
distance and expansion rate, even without a prediction for
the expansion rate.
(If backreaction is significant, the expansion rate is not expected
to be the same as in the \LCDM model, so increasing agreement of
observations with the \LCDM model also tests backreaction indirectly.)
In a FRW universe there is a specific relation between distance
and expansion rate, which can be used to test the FRW metric
\cite{Clarkson:2007b}.
If backreaction is significant, the distance--expansion rate relation
is in general different from the FRW case \cite{Rasanen:2008b, Rasanen:2009b}.
However, so far there has been no reliable estimate of the expected
magnitude of the deviation in the real universe, nor of its redshift
dependence.
(A calculation with an ad hoc treatment of redshift and distance
not based on light propagation equations was presented in
\cite{Larena:2008}; see also \cite{Rasanen:2007}.)

In this paper we consider a backreaction toy model where the
expansion rate has features similar to those expected if backreaction
is significant in the real universe.
First, we show that the model fits the Union2.1 SN data
\cite{Suzuki:2011} well, even though the expansion history is
quite different from the \LCDM case.
Second, we quantify the violation of the FRW consistency
condition between distance and expansion rate.

In \sec{sec:main} we present the toy model, the
results of the fit to the Union2.1 SN data and discuss
the expansion history.
In \sec{sec:sign} we consider effective equations of state, the
deviation from the FRW relation between distance and expansion rate,
and cosmological observations other than SNe.
We conclude in \sec{sec:conc}.

\section{Backreaction toy model and datafit} \label{sec:main}

\subsection{Backreaction toy model}

\para{The expansion rate.}

In a statistically homogeneous and isotropic universe
with a finite homogeneity scale, the average expansion rate
describes the overall expansion of any region larger than the
homogeneity scale. If the distribution of structures evolves
slowly compared to the time it takes for light to cross the
homogeneity scale, redshift and distance are
expected to be calculable from the average expansion rate
\cite{acc, peak, Rasanen:2008b, Rasanen:2009b, Bull:2012}.
The homogeneity scale in the real universe appears to be of the
order 100 Mpc, although it is debated whether this has been
satisfactorily established observationally
\cite{Hogg:2005, inhom, Scrimgeour:2012, Nadathur:2013}.
In any case, average quantities are not expected to give
an accurate description on much smaller scales.

If backreaction increases the expansion rate significantly,
this is related to large variance of the expansion rate due
to increasing differentiation of overdense and underdense
regions in the course of structure formation.
We consider a simple toy model that allows for such evolution.
The expansion rate is taken to
be that of a union of two FRW dust models, one with
positive and the other with negative spatial curvature.
We take the dust energy density to be always non-negative.
This is similar to the toy model studied in \cite{acc},
but there the negatively curved model was taken to be completely empty.
The universe is not considered to be physically composed of two regions,
this is just a simple way to parametrise an expansion rate that
in a realistic situation would arise from an average over
a complicated distribution of structures.
The positively curved part describes overdense regions,
which slow down relative to the spatially flat case, and the negatively
curved part represents underdense regions, which speed up relative
to the spatially flat model.

The volume of the universe is proportional to $a^3=a_1^3+a_2^3$,
where $a_1$ and $a_2$ are the scale factor of a negatively and positively
curved FRW dust model, respectively. The average expansion rate is
\bea \label{HBR}
  H \equiv \H &=& \frac{ a_1^3 }{ a_1^3 + a_2^3 } H_1 + \frac{ a_2^3 }{ a_1^3 + a_2^3 } H_2 \equiv v_1 H_1 + v_2 H_2 \ ,
\eea

\noindent where dot stands for derivative with respect to time $t$
(physically, this is the proper time of observers comoving
with the hypersurface of statistical homogeneity and isotropy
\cite{acc, peak, Rasanen:2008b, Rasanen:2009b}) and $H_i\equiv\adot_i/a_i$.
The average energy density of a dust universe is $\av{\rhom}\propto a^{-3}$.

The scale factors and the time $t$ are given in terms
of the development angle $\phi$ and the evolution variable $\phit$ as
\bea
  \label{r1} a_1 &=& A_1 ( \cosh\phit - 1 ) \ , \qquad t = T_1 ( \sinh\phit - \phit ) \\
  \label{r2} a_2 &=& A_2 ( 1 - \cos\phi ) \ , \qquad\ t = T_2 ( \phi - \sin\phi ) \ ,
\eea

\noindent where $A_i$ and $T_i$ are constants. The development
angle varies from $0$ to $2\pi$ and $\phit$ runs from $0$
to $\infty$. The big bang is at $\phi=0$. The function
$\phit(\phi)$ is determined by equating $t$ in \re{r1} and \re{r2}.
At early times, both regions behave like the Einstein--de Sitter (EdS)
universe, the spatially flat matter-dominated FRW model.
The underdense region decelerates less as time goes on, and
asymptotically approaches the coasting empty universe.
The overdense region slows down more over time, turns around 
from expansion to collapse at $\phi=\pi$ and collapses to a
singularity at $\phi=2\pi$.
The toy model shows different behaviour for different parameter values.
The universe can transition smoothly from early EdS
behaviour towards a negatively curved FRW model, or there can be a
period of extra deceleration followed by acceleration \cite{acc}.
Note that acceleration may not be needed to explain the observations,
because the relation between expansion rate and distance is different
than in the FRW case \cite{Rasanen:2008b, Rasanen:2009b}.

Physics depends on $A_1$ and $A_2$ only via the ratio $A_1/A_2$
and dimensionless quantities depend on $T_1$ and $T_2$ only via $T_1/T_2$.
A large value of $A_1/A_2$ corresponds to large relative volume of
the underdense region, and a large value of $T_1/T_2$ means that
the underdense region evolves rapidly compared to the overdense region.
It is more convenient to have parameters with a clearer physical
interpretation and finite range. Therefore, instead of $A_i$ and $T_i$,
we parametrise the model with the fraction of volume occupied by the
overdense region in the beginning and at the time when it starts to collapse,
\bea \label{f}
  f_\mathrm{b} &\equiv& v_2(\phi=0) = \frac{A_2^3}{A_1^3 (T_2/T_1)^2 + A_2^3} \el
  f_\mathrm{c} &\equiv& v_2(\phi=\pi) = \frac{8 A_2^3}{A_1^3 [\cosh\phit(\pi)-1]^3 + 8 A_2^3} \ .
\eea

\noindent The expansion history of the universe is determined by
the constants $f_\mathrm{b}$ and $f_\mathrm{c}$, which have the
ranges $0<f_\mathrm{b}<1$,
\mbox{$0<f_\mathrm{c}<[\frac{9\pi^2}{16}(f_\mathrm{b}^{-1}-1)+1]^{-1}$}.
In addition, there is one parameter that determines at which point
of the expansion history the universe is today.
We take this to be the value of the development angle today,
$\phi_0$ (the subscript 0 refers to quantities evaluated today).
Were we to consider dimensional quantities, we would also have
the value of the Hubble parameter today, $H_0$.

For comparison, in the $\Lambda$CDM model, where the energy density
consists of dust and vacuum energy, there is only one free parameter
(we take the name \LCDM to imply spatial flatness), which determines
at which point of the expansion history the universe is today.
(Again excepting the overall scale, usually given by $H_0$.)
We take this to be the value of the matter density parameter
$\Om\equiv8\pi\GN\rhom/(3 H^2)$ (where $\GN$ is Newton's constant)
today, so we have
\bea \label{LCDM}
  H = H_0 \sqrt{ 1 - \Omn + \Omn a^{-3} } \ ,
\eea

\noindent where we have adopted the normalisation $a_0=1$.
We also consider the OCDM model, which is the FRW model
where the energy density consists of dust and the universe
has negative spatial curvature. The expansion rate of the
OCDM model is
\bea \label{OCDM}
  H = H_0 \sqrt{ (1 - \Omn) a^{-2} + \Omn a^{-3} } \ .
\eea

\para{The redshift and the distance.}

We take the redshift $z$ to be given by \cite{Rasanen:2008b, Rasanen:2009b}
\bea \label{z}
  1 + z &=& a^{-1} \ .
\eea

\noindent We assume that the angular diameter distance $D_A$
can be solved from \cite{Rasanen:2008b, Rasanen:2009b}
\bea \label{DA}
  H \frac{\rmd}{\rmd z} \left[ (1+z)^2 H \frac{\rmd D_A}{\rmd z} \right] &=& - 4\pi\GN\av{\rhom} D_A
\eea

\noindent with the initial conditions
$D_A(0)=0$, $\frac{\rmd D_A}{\rmd z}(0)=H_0^{-1}$.
In the FRW case, $\av{\rhom}$ would be replaced by $\rho+p$,
where $p$ is the pressure. Given $\av{\rhom}\propto a^{-3}=(1+z)^3$,
\re{DA} can be written as
\bea \label{dA}
  h \frac{\rmd}{\rmd z} \left[ (1+z)^2 h \frac{\rmd d_A}{\rmd z} \right] &=& - \frac{3}{2} \Omn (1+z)^3 d_A \ ,
\eea

\noindent where we have introduced the notation $h\equiv H/H_0$,
$d_A\equiv H_0 D_A$; we also denote $d\equiv (1+z) d_A$.
The distance is fully determined by the expansion history
$h(z)$ and the value of $\Omn$. The luminosity distance is
$D_L=(1+z)^2 D_A$ \cite{Etherington:1933, Ellis:1971};
we denote $d_L\equiv H_0 D_L$.

\subsection{Fit to Union2.1 data} \label{sec:data}

\para{Supernova data fit.}

We fit the backreaction toy model defined by \re{HBR}--\re{f}
to the Union2.1 SN data \cite{Suzuki:2011} using
a Monte Carlo Markov Chain analysis.
For comparison, we also fit the \LCDM model \re{LCDM} and
the OCDM model \re{OCDM} to the same data.
The OCDM model is included to illustrate how much backreaction
improves the fit compared to the FRW dust case.
The Union2.1 dataset consists of distances to 580
type Ia SNe between redshifts 0.015 and 1.4.
Distances of the Union SNe are determined only relative to
each other, the absolute distance scale is arbitrary, so
the data only fixes $d_L$, not $D_L$.

The analysis leading to the publicly available Union2.1
distance data has been done assuming that
the \LCDM model is correct. As the SN light curve
parameters and cosmological parameters have been fitted at
the same time, the resulting distances are model-dependent,
and may be biased towards the \LCDM model.
To properly consider other models, the data should be re-analysed
from the beginning \cite{Nadathur:2010}.
The results also depend on the choice of light curve fitter \cite{fitter}.
Furthermore, as the intrinsic dispersion of SN luminosity
is adjusted to obtain a good fit with the \LCDM model,
absolute goodness-of-fit values are not meaningful.
Nevertheless, relative goodness-of-fit statistics can be
compared, bearing the above caveats in mind.

If the goal were to confront the result of a realistic
calculation of the expansion rate due to backreaction with
observations, it would be important to address the above issues.
However, we are interested in seeing whether a toy model
with realistic features can provide a reasonable fit to the data
and in estimating the deviation of the distance--expansion rate
relation from the FRW case, so the precise quality of the fit
and the exact parameter values are not crucial. Likewise, the fact
that the toy model has two more parameters than the \LCDM model
is not important.

\para{Results of the data fit.}

We highlight two sets of parameter values, called models 1 and 2.
Model 1 is the best-fitting model we found that satisfies
the constraint $\Omn\leq0.35$. In order to demonstrate a
different backreaction expansion history with a smaller
matter density parameter, we also show model 2.
The parameter values and $\chi^2$ numbers for models 1 and
2, as well as for the best-fit \LCDM and OCDM models,
are listed in table \ref{tab:paras}. We also give the values of $\Omn$ and
$q_0$, where $q$ is the deceleration parameter, defined in \re{q}.

\begin{table}
\begin{center}
\begin{tabular}{|c|ccccccc|c|}
\hline
Model & $\chi^2$ & $\chi^2/\textrm{d.o.f.}$ & $\Omn$ & $q_0$ & $f_\mathrm{b}$ & $f_\mathrm{c}$ & $\phi_0$ \\
\hline\hline
Model 1 & 548 & 0.94 & 0.35 & -0.52 & 0.995 & 0.215 & 6.24 \\
Model 2 & 551 & 0.95 & 0.28 & -0.39 & 0.974 & 0.168 & 5.99  \\
\hline
\LCDM   & 545 & 0.94 & 0.29 & -0.57 & -     & -     & - \\
OCDM    & 566 & 0.98 & 0.00 & 0.00  & -     & -     & - \\
\hline
\end{tabular}
\end{center}
\caption{Parameter values for backreaction models 1 and 2 and the
best-fit \LCDM and OCDM models.}
\label{tab:paras}
\end{table}

The backreaction models are a significant improvement over the best-fit
OCDM model (which is, as expected, the limit of a completely
empty universe), and the quality of the fit is close to the \LCDM
model. The $\chi^2$/d.o.f. is 0.94 for both the \LCDM model and
model 1, and 0.95 for model 2 (note that the data points
are not independent). Taking into account the issues
discussed above, the difference in $\chi^2$ is not large,
$\Delta\chi^2=3$ for model 1 and $\Delta\chi^2=5$ for model 2.
In both backreaction models, $f_\mathrm{b}$ is close to 1, and
it is possible that a more detailed scan of the parameter space
in the region around $f_\mathrm{b}=1$ would uncover models
that fit the data better. Relaxing the constraint $\Omn\leq0.35$
does lead to better fitting models, with expansion histories even
further away from the \LCDM model. As already mentioned, we are
more interested in demonstrating qualitative features than in
finding the optimal fit.

\begin{figure}
\begin{minipage}[t]{7.7cm} 
\scalebox{1.0}{\includegraphics[angle=0, clip=true, trim=0.8cm 0.6cm 1.5cm 1.5cm, width=\textwidth]{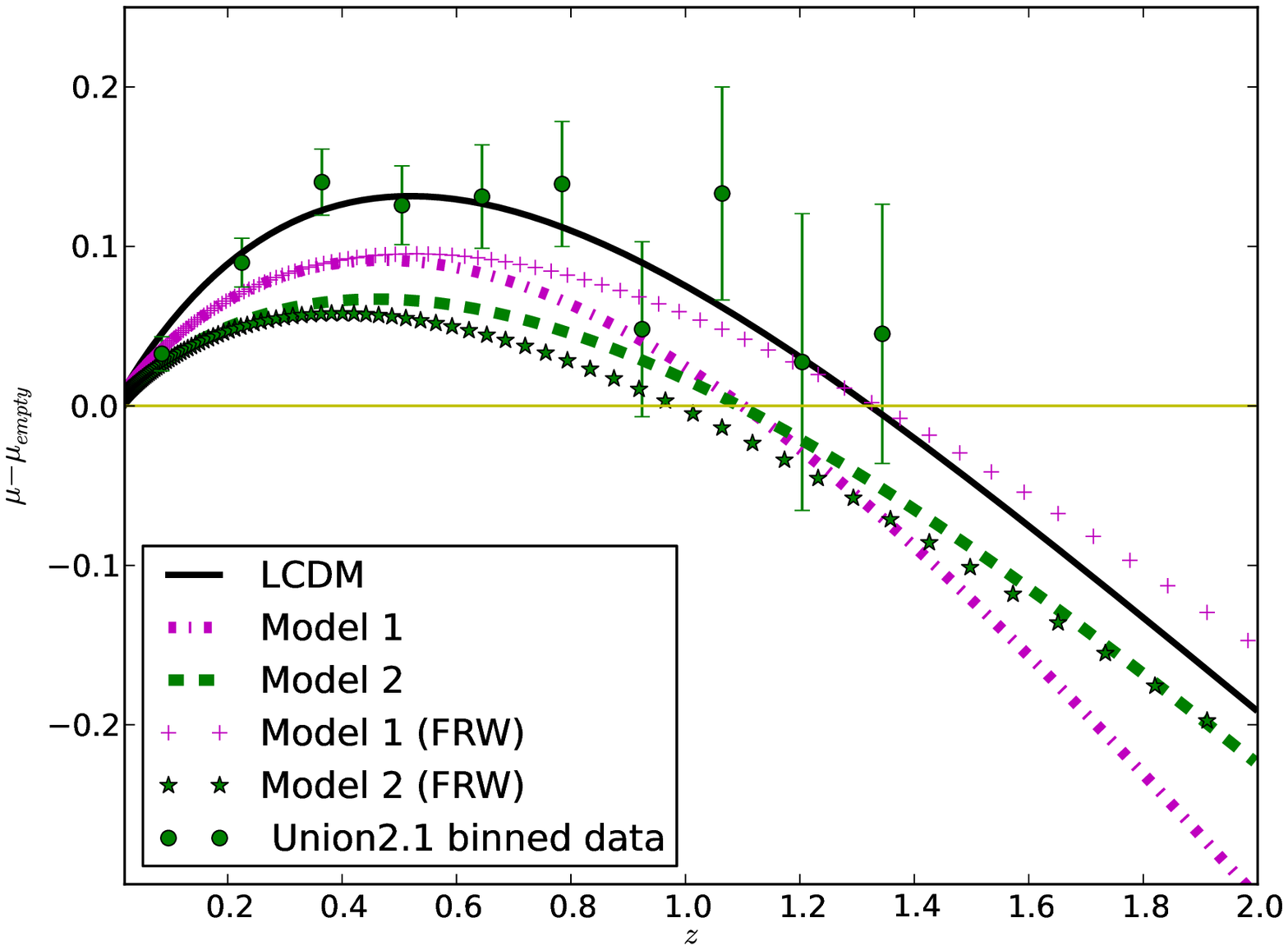}}
\begin{center} {\bf (a)} \end{center}
\end{minipage}
\begin{minipage}[t]{7.7cm}
\scalebox{1.0}{\includegraphics[angle=0, clip=true, trim=0.6cm 0.6cm 1.7cm 1.5cm, width=\textwidth]{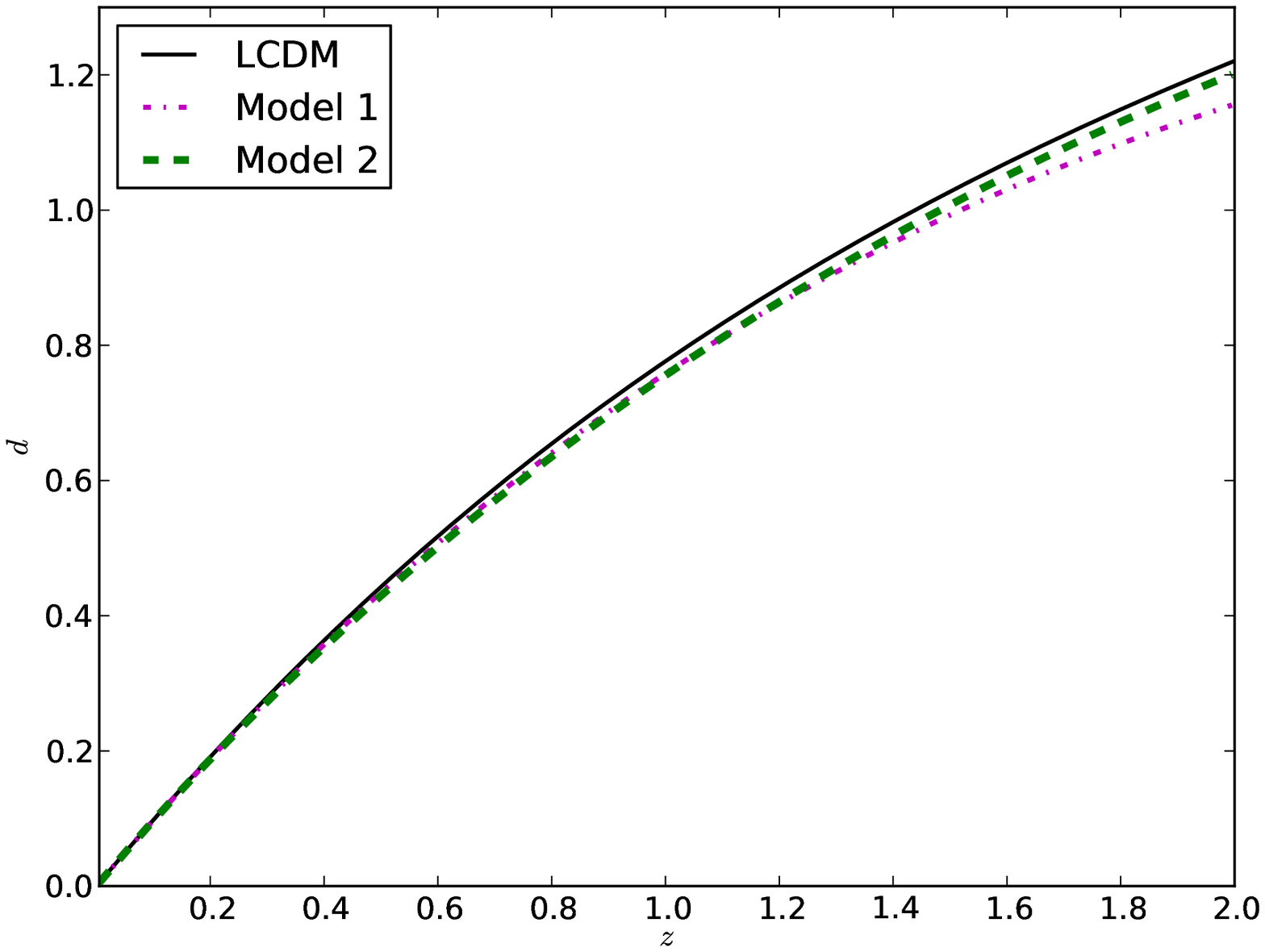}}
\begin{center} {\bf (b)} \end{center}
\end{minipage}
\caption{a) Magnitude relative to the empty FRW universe, corresponding
to the parameters given in table \ref{tab:paras}. Dots indicate
binned Union2.1 data. Crosses and stars mark the magnitude
that models 1 and 2 would have were the relation between distance
and expansion rate the same as in the FRW case; see \sec{sec:dev}.
b) The distance $d=(1+z) d_A$.}
\label{fig:mud}
\end{figure}

In \fig{fig:mud}a we show\footnote{Plots are made with matplotlib
\cite{Hunter:2007}.} the magnitude--redshift relation in models 1 and 2
and in the \LCDM model, relative to the empty FRW universe,
$\mu-\mu_{empty}=5\log_{10}(d/d_{empty})$.
The corresponding distances are shown in \fig{fig:mud}b.
The fact that the curves for the backreaction models are below the
\LCDM line at small $z$ reflects the fact that they accelerate
less strongly today, but this is not a general feature.
There are parameter values that give a good fit with
more rapid acceleration today. In both model 1 and 2,
$\phi_0$ is not far from the maximum value $2\pi$, indicating
that the overdense region is rapidly collapsing
and its fraction of the volume is small.
The acceleration results from the universe being first dominated
by the slowly expanding region, until it is overtaken by the faster
expanding underdense region. The more rapid the transition, the faster
the acceleration. In the future, the backreaction models will return
to deceleration, asymptotically approaching the empty universe.

\begin{figure}
\begin{minipage}[t]{7.7cm} 
\scalebox{1.0}{\includegraphics[angle=0, clip=true, trim=1.2cm 0.3cm 1.2cm 1.5cm, width=\textwidth]{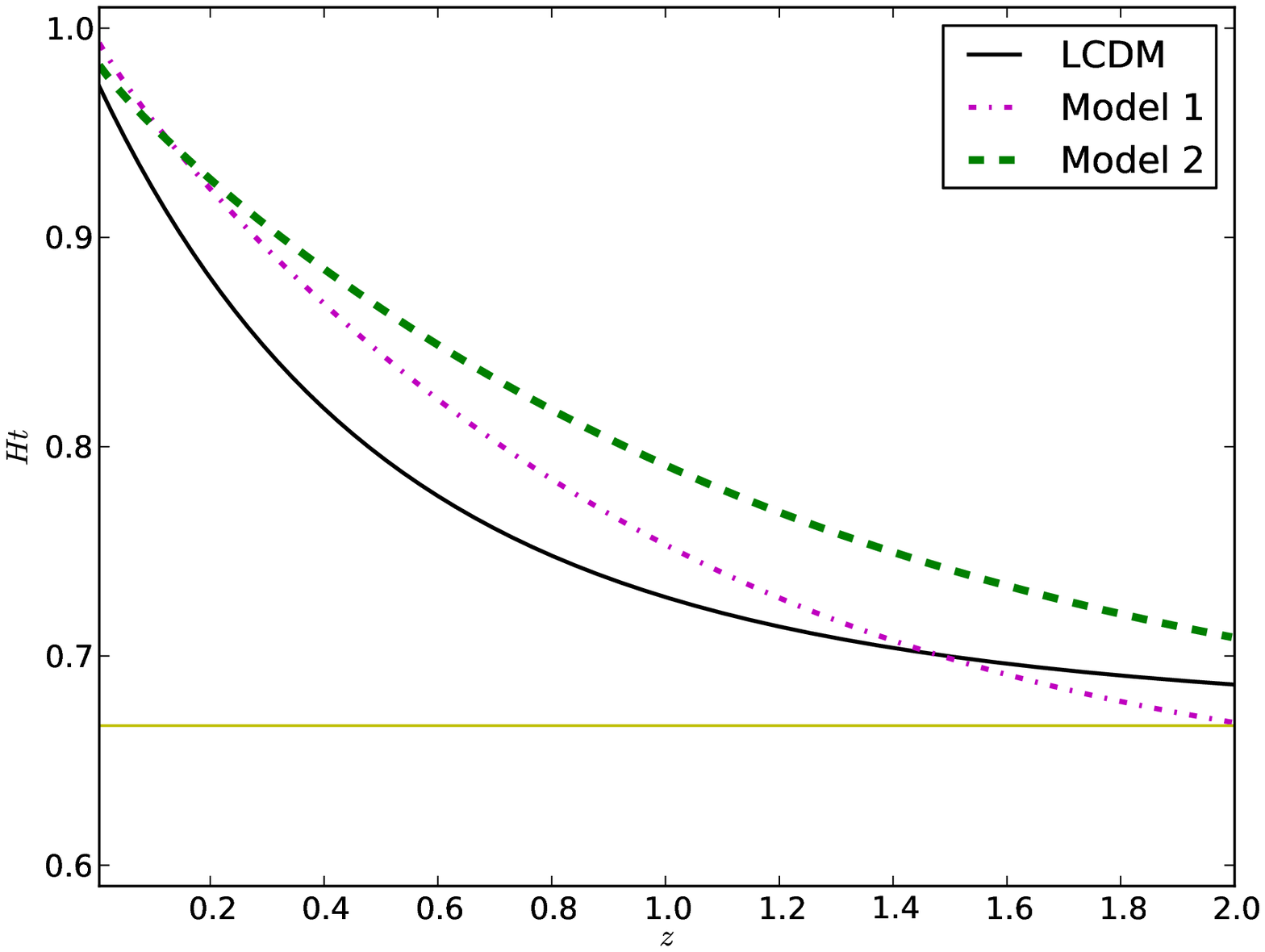}}
\begin{center} {\bf (a)} \end{center}
\end{minipage}
\begin{minipage}[t]{7.7cm}
\scalebox{1.0}{\includegraphics[angle=0, clip=true, trim=0.6cm 0.3cm 1.7cm 1.5cm, width=\textwidth]{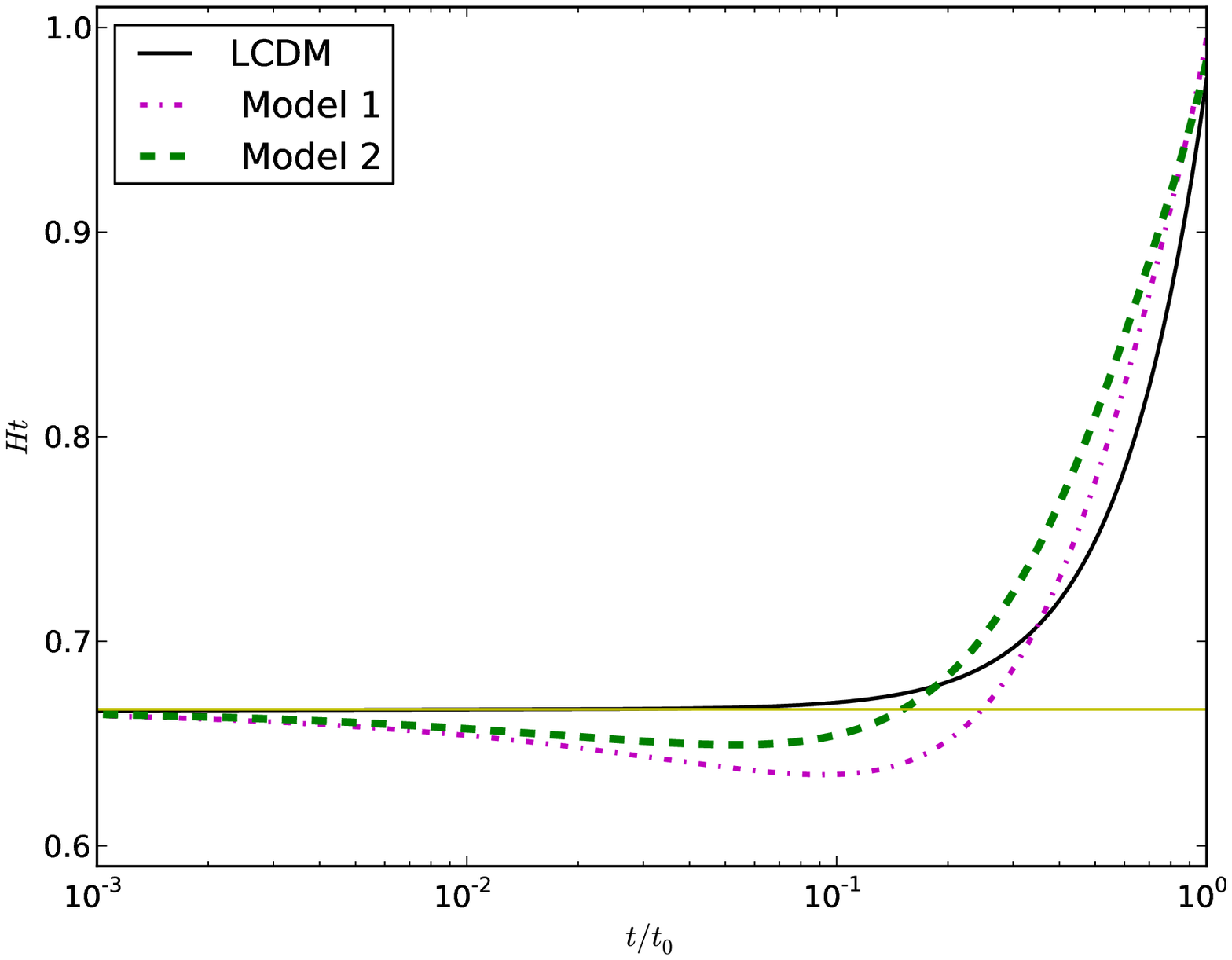}}
\begin{center} {\bf (b)} \end{center}
\end{minipage}
\caption{a) The Hubble parameter times the age of the universe as function
of $z$. The horizontal line marks the EdS value $Ht=\frac{2}{3}$.
b) The same quantity as function of $t/t_0$. The value $t/t_0=10^{-3}$
corresponds to $z\approx100$.}
\label{fig:Ht}
\end{figure}

The expansion history is illustrated in \fig{fig:Ht}
in terms of $Ht$. In the \LCDM model, $Ht$ grows
monotonically from $\frac{2}{3}$ at early times to $Ht=0.98$
today, and in the future $Ht$ will increase linearly with $t$.
At sufficiently early times, backreaction models also have $Ht=\frac{2}{3}$.
The expansion slows down as the overdense region becomes important,
and $Ht$ falls below $\frac{2}{3}$. The expansion speeds up
when the underdense region takes over; model 1 has $H_0 t_0=0.995$
and model 2 has $H_0 t_0=0.98$. The extra deceleration (compared
to the EdS case) before the underdense region takes over
makes the change in the expansion rate more drastic,
and thus contributes to stronger acceleration.
It can also lead to a large value of $\Omn$: in model 2, there
is less early deceleration, and correspondingly $\Omn$ is smaller
and the acceleration is weaker than in model 1.
This kind of expansion history is quite different from the \LCDM case,
but it could be mimicked by a FRW model with suitably exotic matter.
We now turn to aspects of backreaction which cannot be
reproduced by any FRW model.

\section{Signatures of backreaction} \label{sec:sign}

\subsection{Deviation from FRW behaviour} \label{sec:dev}

\para{The density parameters.}

To understand the evolution of the backreaction models, it
is useful to consider the Buchert equations, which describe how
the average expansion rate of an irrotational dust universe
is related to the variance of the expansion rate, the shear and
the spatial curvature \cite{Buchert:1999a}:
\bea
  \label{Ray} 3 \frac{\addot}{a} &=& - 4 \pi\GN \av{\rho} + \sQ \\
  \label{Ham} 3 \HH &=& 8 \pi \GN \av{\rho} - \frac{1}{2}\av{\sR} - \frac{1}{2}\sQ \\
  \label{cons} 0 &=& \pat_t \av{\rho} + 3 \H \av{\rho} \ ,
\eea

\noindent where
$\sQ \equiv \frac{2}{3}\left( \av{\theta^2} - \av{\theta}^2 \right) - 2 \av{\sigma^2}$
is the backreaction variable, $\av{}$ stands for spatial
average, $\sigma^2$ is the shear scalar and $\sR$ is the
spatial curvature. Dividing \re{Ray} and \re{Ham} by $3 H^2$,
we get \cite{Buchert:1999a}
\bea \label{omegas}
  \label{q} q &\equiv& - \frac{1}{H^2} \frac{\addot}{a} = \frac{1}{2} \Om + 2 \OQ \\
  \label{Omegas} 1 &=& \Om + \OR + \OQ \ ,
\eea

\noindent where $\Om\equiv 8\pi\GN\av{\rho}/(3 H^2)$,
$\OR\equiv-\av{\sR}/(6H^2)$ and $\OQ\equiv-\sQ/(6H^2)$ are
the density parameters of matter, spatial curvature and
the backreaction variable, respectively.
In addition to the deceleration parameter $q$,
we will use the jerk parameter
$j\equiv\frac{1}{H^3} \frac{1}{a} \frac{\rmd^3 a}{\rmd t^3}$.

\begin{figure}
\begin{minipage}[t]{7.7cm} 
\scalebox{1.0}{\includegraphics[angle=0, clip=true, trim=0.8cm 0.7cm 1.3cm 1.3cm, width=\textwidth]{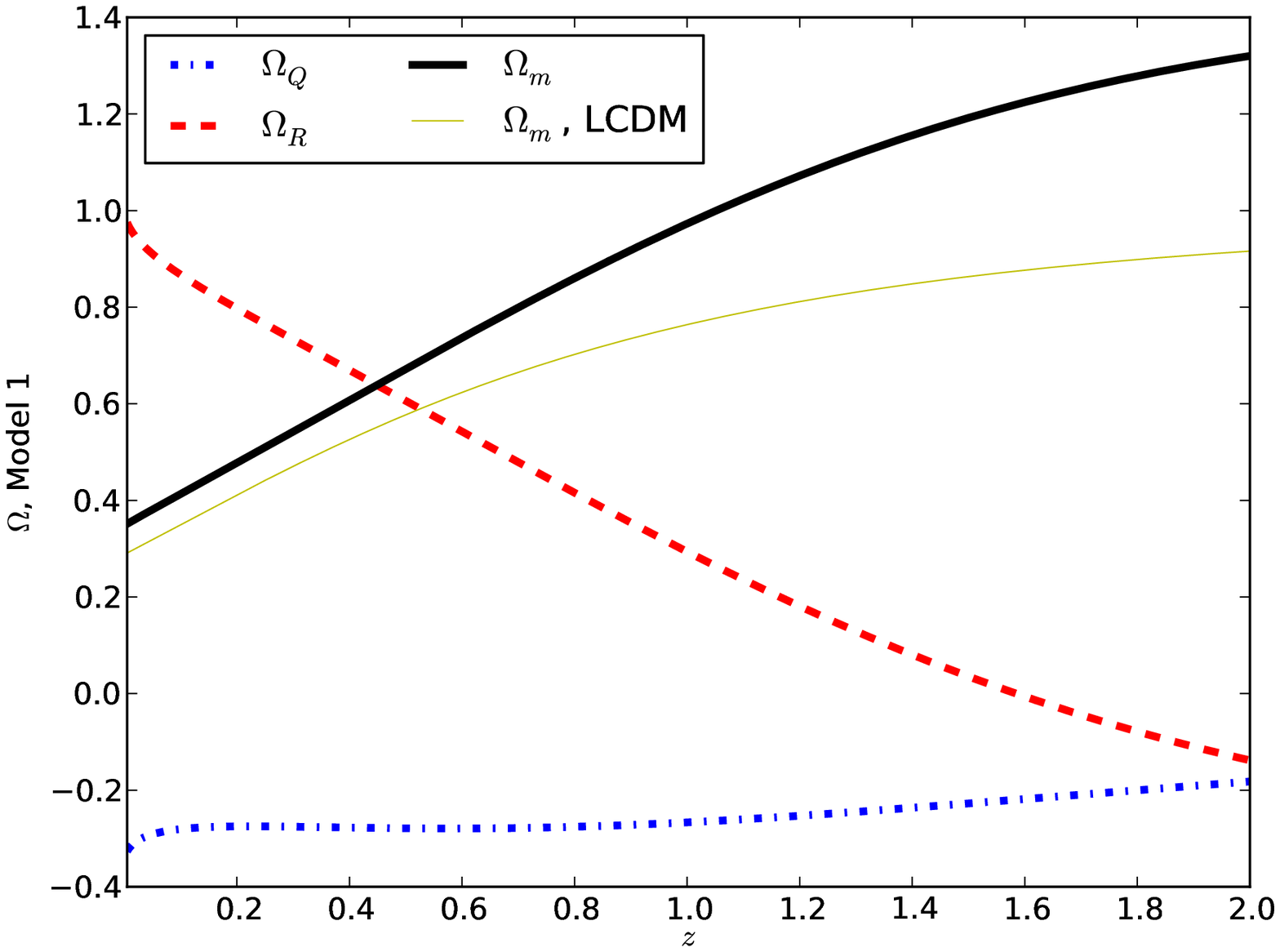}}
\begin{center} {\bf (a)} \end{center}
\end{minipage}
\begin{minipage}[t]{7.7cm}
\scalebox{1.0}{\includegraphics[angle=0, clip=true, trim=0.5cm 0.7cm 1.7cm 1.3cm, width=\textwidth]{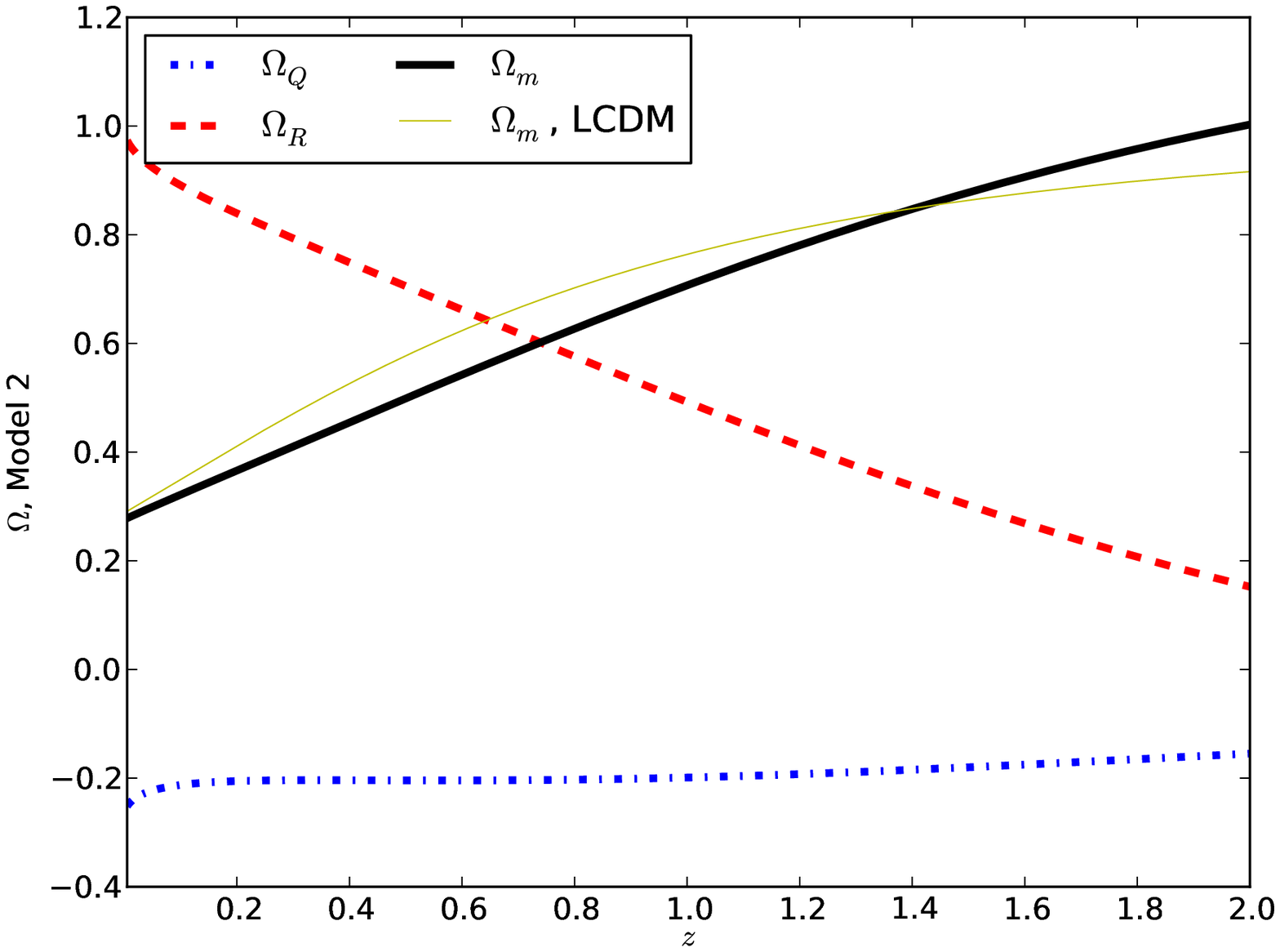}}
\begin{center} {\bf (b)} \end{center}
\end{minipage}
\caption{a) The density parameters \re{Omegas} for a) model 1 and
b) model 2, with $\Om$ for the \LCDM model also plotted.}
\label{fig:Omegas}
\end{figure}

The density parameters as a function of redshift are shown in
\fig{fig:Omegas}. Extra deceleration in model 1 is related to
the fact that $\Om>1$ for $z>1.1$. Spatial curvature is
correspondingly positive for $z>1.6$, but switches sign
as the universe becomes dominated by the underdense region,
and the curvature is large today, $\ORn=1.00$. The backreaction
variable is also large, $\OQn=-0.35$, as required for acceleration.
In model 2, the transition is smoother. The spatial curvature is
always negative, and $\Om$ crosses unity earlier, at $z=2.0$.

\para{FRW consistency condition.}

The backreaction relation \re{DA} between distance and expansion
rate is different than the FRW relation.
The difference can be quantified by a ``fitting model''
expansion rate $\hfit(z)$, defined as the expansion rate of
the spatially flat FRW model with the same distance--redshift
relation as the backreaction model.
In any FRW model where the expansion rate never passes through
zero\footnote{If the expansion rate is not a strictly
monotonic function of time, the redshift is not a valid
time coordinate everywhere. We assume $h>0$,
though $h$ could be allowed to be zero at isolated points.},
distance and expansion rate are related as
\bea \label{dAFRW}
  d(z) = \frac{1}{\sqrt{\OKn}} \sinh\left(\sqrt{\OKn}\int_0^z \frac{\rmd \tilde z}{h(\tilde z)} \right) \ ,
\eea

\noindent where $\OKn$ is the present value of the spatial curvature
density parameter $\OR$. (In the FRW case, it is conventional to denote
the spatial curvature density parameter by $\OK$ instead of $\OR$.)
The spatial curvature of a FRW universe is $\sR=6 K/a^2$, where $K$
is a constant, so $\OK\equiv\OR=-K/(aH)^2$. In the spatially flat case, we have
\bea \label{d}
  d(z) = \int_0^z \frac{\rmd \tilde z }{h(\tilde z)} \ .
\eea

\noindent The fitting model expansion rate is thus
\bea \label{hfit}
  \hfit(z) \equiv \frac{1}{d'(z)} \ ,
\eea

\noindent where prime denotes derivative with respect to $z$.
The fitting model matter density parameter is correspondingly
\mbox{$\Omfit(z)\equiv\Omn (1+z)^3/\hfit(z)^2$}.

The relation \re{dAFRW} can be rewritten as \cite{Clarkson:2007b}
\bea \label{k}
  k &=& \frac{1 - h^2 d'^2}{d^2} \ ,
\eea

\noindent where we have denoted $k\equiv-\OKn=K/H_0^2$.
In the FRW case, $k$ is constant and related to the spatial
curvature. However, \re{k} can be taken as the definition of
$k$ for any set of observations $h(z)$ and $d(z)$.
(We neglect angular variation; in the statistically homogeneous
and isotropic case, this is expected to be small for distances
larger than the homogeneity scale \cite{Rasanen:2008b, Rasanen:2009b}.)
If the expansion rate and light propagation in the universe
are not on average described by a four-dimensional FRW model,
$k$ in general varies with redshift \cite{Ferrer, February:2009},
and it is not necessarily straightforwardly related to the spatial
curvature. The constancy of $k$ was suggested in \cite{Clarkson:2007b}
as a consistency condition for the FRW metric, and the following
quantity was introduced to quantify the deviation of $k$ from constant:
\bea \label{C}
  \mC(z) \equiv - \frac{d^3}{2 d'} k' = 1 + h^2 ( d d'' - d'^2 ) + h h' d d' \ .
\eea

\noindent A non-zero value of $\mC$ at any $z$ indicates that
the universe cannot be described by a four-dimensional FRW metric.

\begin{figure}[t]
\begin{minipage}[t]{7.7cm} 
\scalebox{1.0}{\includegraphics[angle=0, clip=true, trim=1.0cm 0.7cm 1.2cm 1.4cm, width=\textwidth]{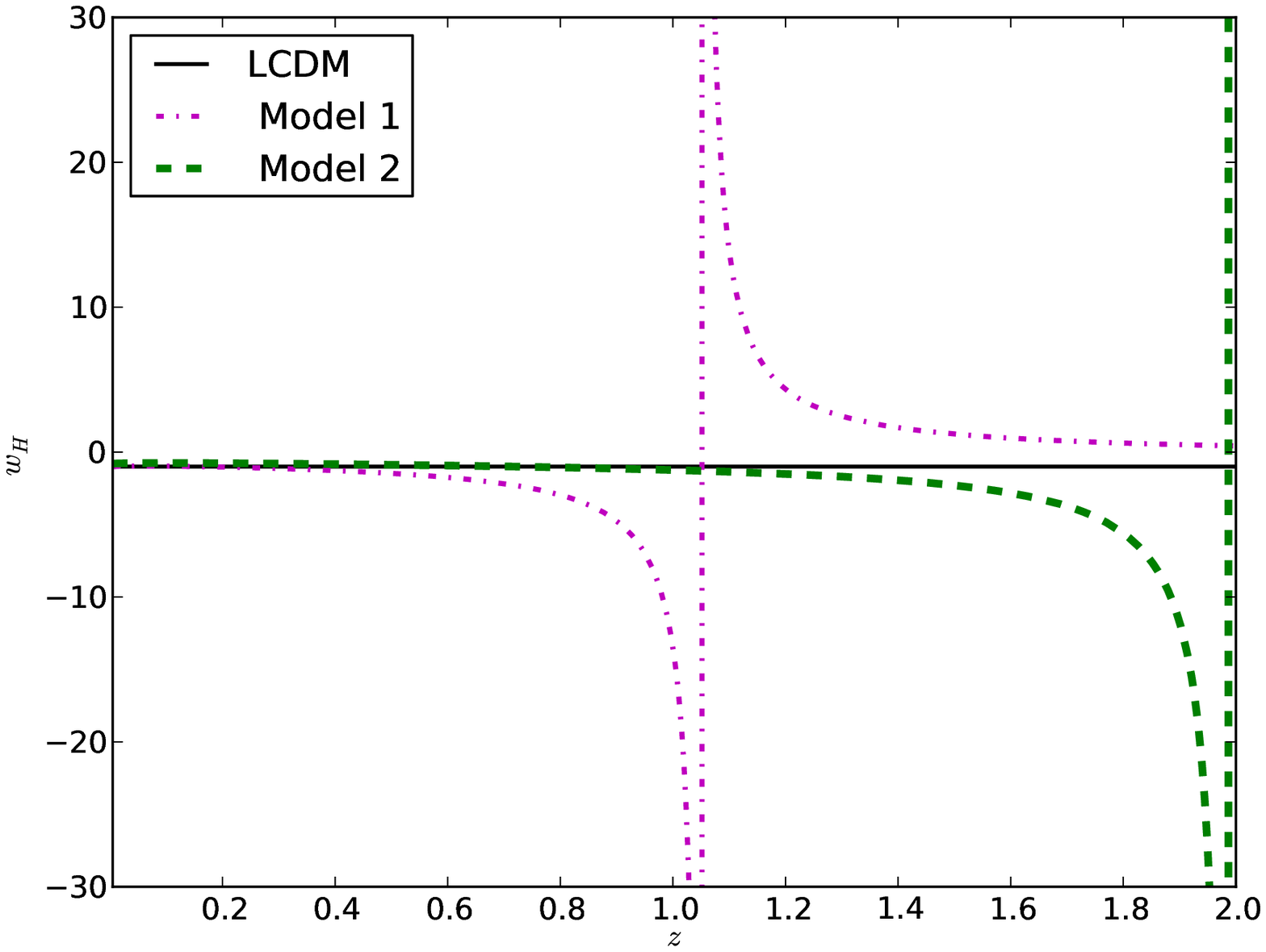}}
\begin{center} {\bf (a)} \end{center}
\end{minipage}
\begin{minipage}[t]{7.7cm}
\scalebox{1.0}{\includegraphics[angle=0, clip=true, trim=0.8cm 0.7cm 1.4cm 1.3cm, width=\textwidth]{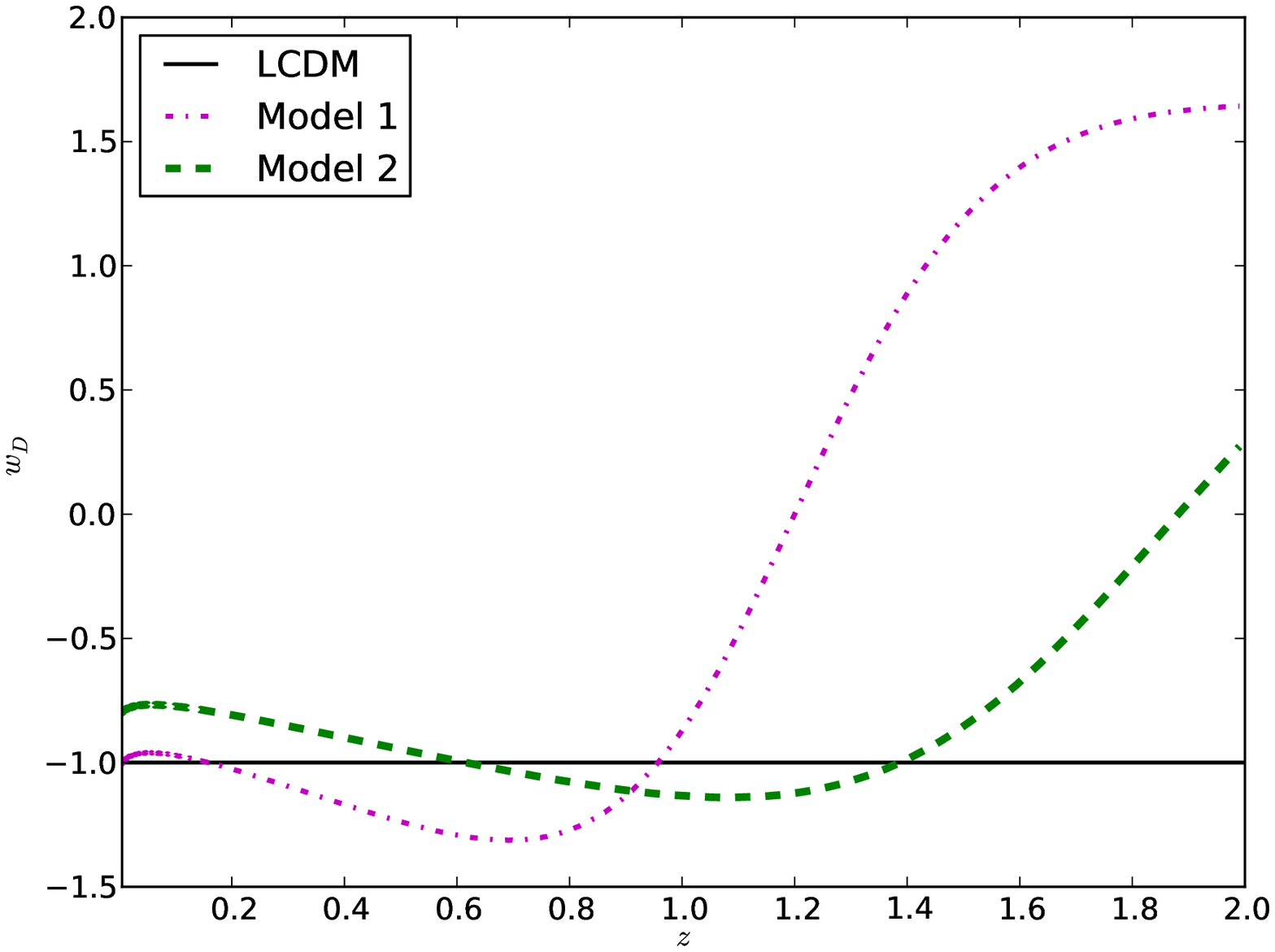}}
\begin{center} {\bf (b)} \end{center}
\end{minipage}
\caption{a) The effective equation of state $w_H$ that corresponds to the expansion rate.
b) The effective equation of state $w_D$ that corresponds to the distance.}
\label{fig:w}
\end{figure}

Using \re{dA}, we can define a quantity analogous to $k$ for backreaction:
\bea \label{m}
  m(z) \equiv - \frac{2 h}{3 (1+z) d} [ h d'' + h' d' - (1+z)^{-1} h' d ] \ .
\eea

\noindent The quantity $m$ is constant and equal to $\Omn$
in backreaction, as well as in FRW models where the energy density
consists of dust and vacuum energy. In all other FRW models,
$m$ depends on $z$. In correspondence with $\mC$, we can also
define $\mB\equiv m'$, so that $\mB$ vanishes for backreaction.
Just as constancy of $k$ (vanishing of $\mC$) is a test of
the FRW metric, constancy of $m$ (vanishing of $\mB$) provides a
test of backreaction. (With the caveat that the relations
\re{z} and \re{DA} for redshift and distance should be placed
on a more rigorous footing.)
At small $z$, $\mC$ and $\mB$ are less prone to observational errors
than $k$ and $m$ in the sense that the latter involve the ratio
of two quantities that both approach zero as $z$ goes to zero.
However, determining derivatives of $d$ is challenging,
so it may be easier to use $k$ and $m$, which involve one less
derivative than $\mC$ and $\mB$, respectively.

\para{Effective equations of state.}

A backreaction model can be equivalently parametrised with
two different effective equations of state, one of which
corresponds to a FRW model with the same expansion history
$h(z)$ and the other to a FRW model with the same
distance--redshift relation $d(z)$.
Specifically, consider a spatially flat FRW model where the energy
density consists of dust (with the same value of $\Omn$ as in
the backreaction model) plus an exotic component.
We define the effective expansion equation of state $w_H(z)$
as the equation of state of the exotic component, assuming that
the FRW model has the same expansion history $h(z)$ as the
backreaction model.
Correspondingly, the effective distance equation of state
$w_D(z)$ is defined as the equation of state for which the FRW
model has the same distance--redshift relation $d(z)$ as
the backreaction model. The effective expansion equation of state is
\bea \label{wH}
  w_H(z) &\equiv& \frac{2 (1+z) \frac{h'}{h} - 3}{3 - 3 \Om} = \frac{2 q - 1}{3 - 3 \Om} = \frac{\Om + 4 \OQ - 1}{3 - 3 \Om} \ ,
\eea

\noindent where we have in the last equality used \re{q}.
In $w_D(z)$, the quantities $h(z)$ and $\Om(z)$ are replaced
by $\hfit(z)$ and $\Omfit(z)$, respectively. (The second and third
equality in \re{wH} only hold for $w_H$.)
The functions $w_H$ and $w_D$ correspond to the equation of state
of a new exotic component that would be inferred from observations
of the expansion rate and distance, respectively, if they are
analysed in the framework of a spatially flat FRW model.
This parametrisation is problematic if the effective component
at different times decreases and increases the expansion rate $h$
or $\hfit$, because in between the effective energy density passes
through zero and the equation of state diverges.
It is therefore more informative to use equations of state that
correspond to the total effective energy density and pressure,
instead of just the exotic component. The effective total expansion
equation of state is
\bea \label{wHtot}
  \wHtot(z) &\equiv& \frac{2}{3} (1+z) \frac{h'}{h} - 1 = \frac{2 q - 1}{3} = \frac{\Om + 4 \OQ - 1}{3} \ ,
\eea

\noindent and $\wDtot$ is correspondingly defined with $h$ replaced
by $\hfit$. (Again, the second and third equality only hold for $\wHtot$.)


\begin{figure}[t]
\begin{minipage}[t]{7.7cm} 
\scalebox{1.0}{\includegraphics[angle=0, clip=true, trim=0.9cm 0.7cm 1.2cm 1.4cm, width=\textwidth]{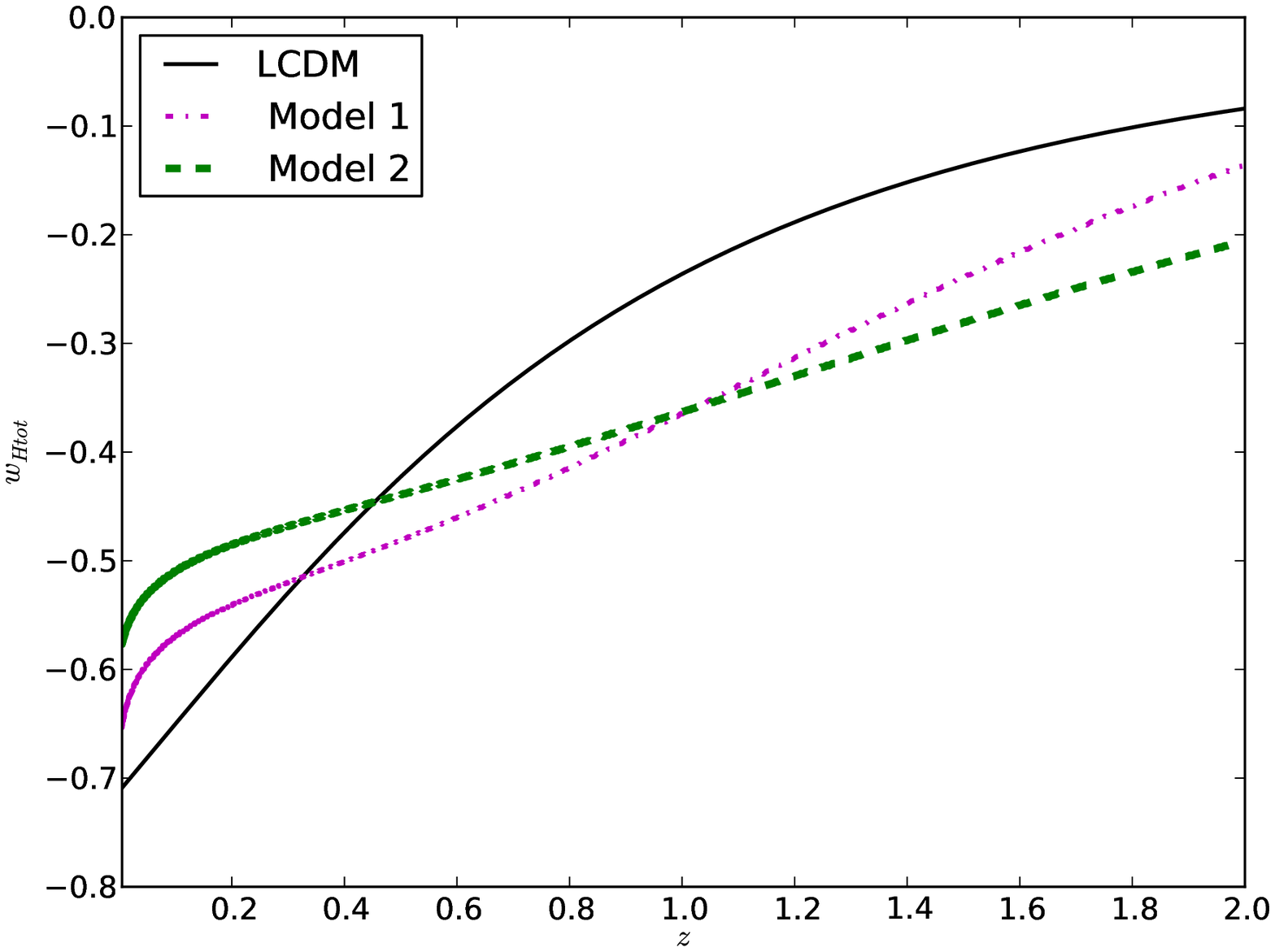}}
\begin{center} {\bf (a)} \end{center}
\end{minipage}
\begin{minipage}[t]{7.7cm}
\scalebox{1.0}{\includegraphics[angle=0, clip=true, trim=0.7cm 0.7cm 1.4cm 1.4cm, width=\textwidth]{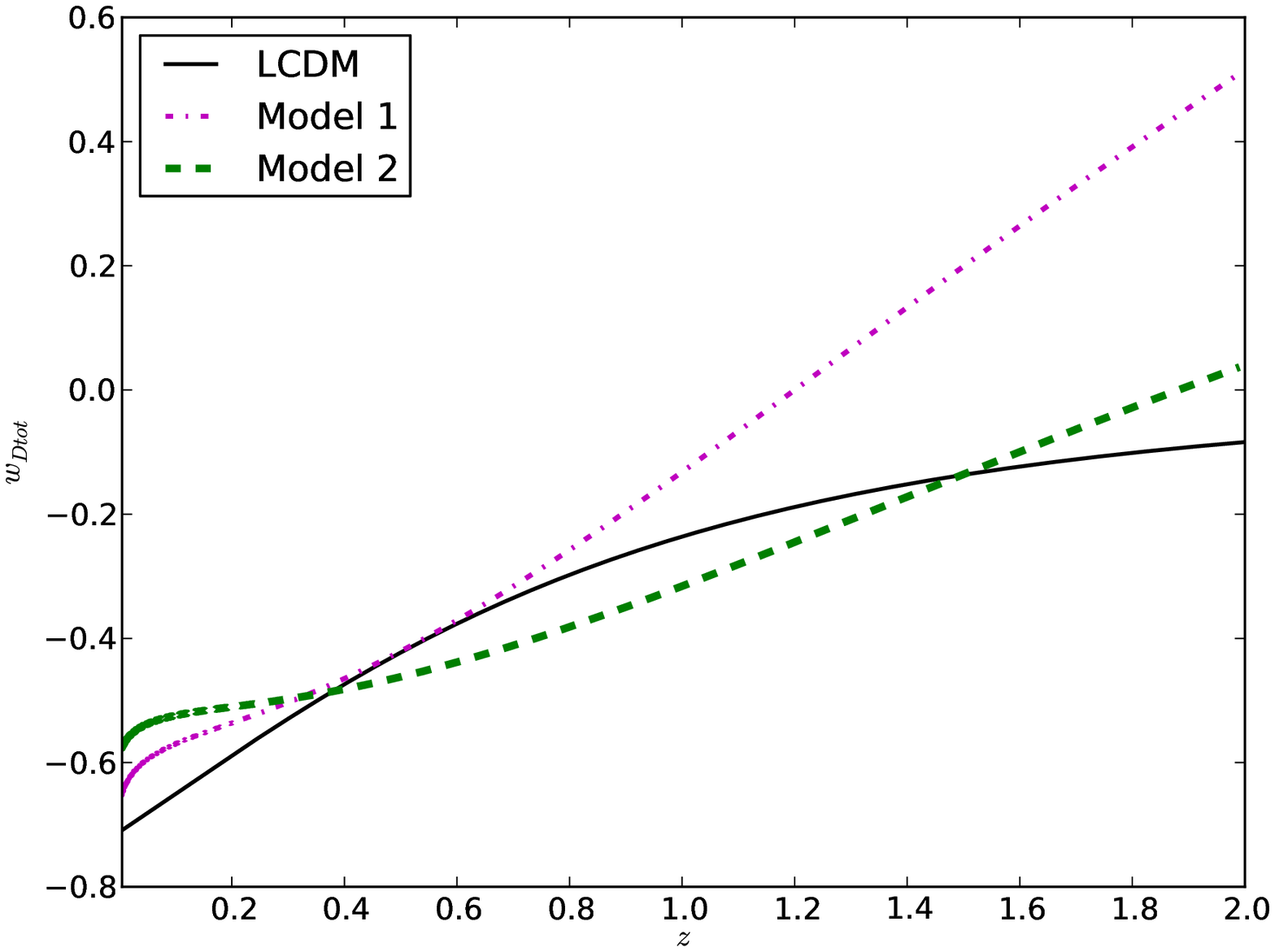}}
\begin{center} {\bf (b)} \end{center}
\end{minipage}
\caption{a) The effective total expansion rate equation of state $\wHtot$.
b) The effective total distance equation of state $\wDtot$.}
\label{fig:wtot}
\end{figure}

In \fig{fig:w} we show the effective equations of state $w_H$ and $w_D$.
The expansion equation of state $w_H$ diverges at the redshift
where $\Om=1$, which is $z=1.1$ for model 1 and $z=2.0$ for model 2.
The distance equation of state $w_D$ is finite, it oscillates
around $-1$ at low $z$ and becomes positive at high $z$.
This smooth behaviour of $w_D$ is not generic.
There are well-fitting models for which
$\Omfit=1$ in the observable range of redshifts, so $w_D$
diverges, demonstrating that an equation of state can be far from
$-1$ and far from constant and yet fit the data well,
underlining model-dependence of limits on the equation of
state and its change with redshift \cite{trans}.

The total equations of state $\wHtot$ and $\wDtot$ shown in
\fig{fig:wtot} give a clearer picture of how the
difference in the distance--expansion rate relation affects
the effective equation of state. For both $\wHtot$ and $\wDtot$,
there are significant differences between the backreaction models
and the \LCDM model, but $\wDtot$ is closer to the \LCDM line,
in particular for model 2. In the backreaction models, $\wHtot$
crosses the value $-\frac{1}{3}$ at an earlier time (a higher redshift)
than in the \LCDM model, so acceleration starts earlier and is
stronger than in the \LCDM model; the trend reverses at small redshifts.
This difference is smaller in $\wDtot$.
The distance depends on the equation of state via a double integral,
so the biasing can be more clearly seen in \fig{fig:mud}a,
where crosses and stars mark the magnitude--redshift relation
that would correspond to $\wHtot$ in the FRW case.
In model 2, the distance is shifted towards the \LCDM
result by the different distance--expansion rate relation,
but this does not appear to be the case for model 1.
(The shifts correspond to changes of $\Delta\chi^2=+1$
for model 1 and $\Delta\chi^2=-2$ for model 2.)
However, to properly evaluate the bias, we should compare
backreaction models and \LCDM models with the same value of $\Omn$,
and for model 1 the value of $\Omn$ is rather different from
its value in \LCDM model shown.

\begin{figure}[t]
\begin{minipage}[t]{7.7cm} 
\scalebox{1.0}{\includegraphics[angle=0, clip=true, trim=0.9cm 0.7cm 1.2cm 1.5cm, width=\textwidth]{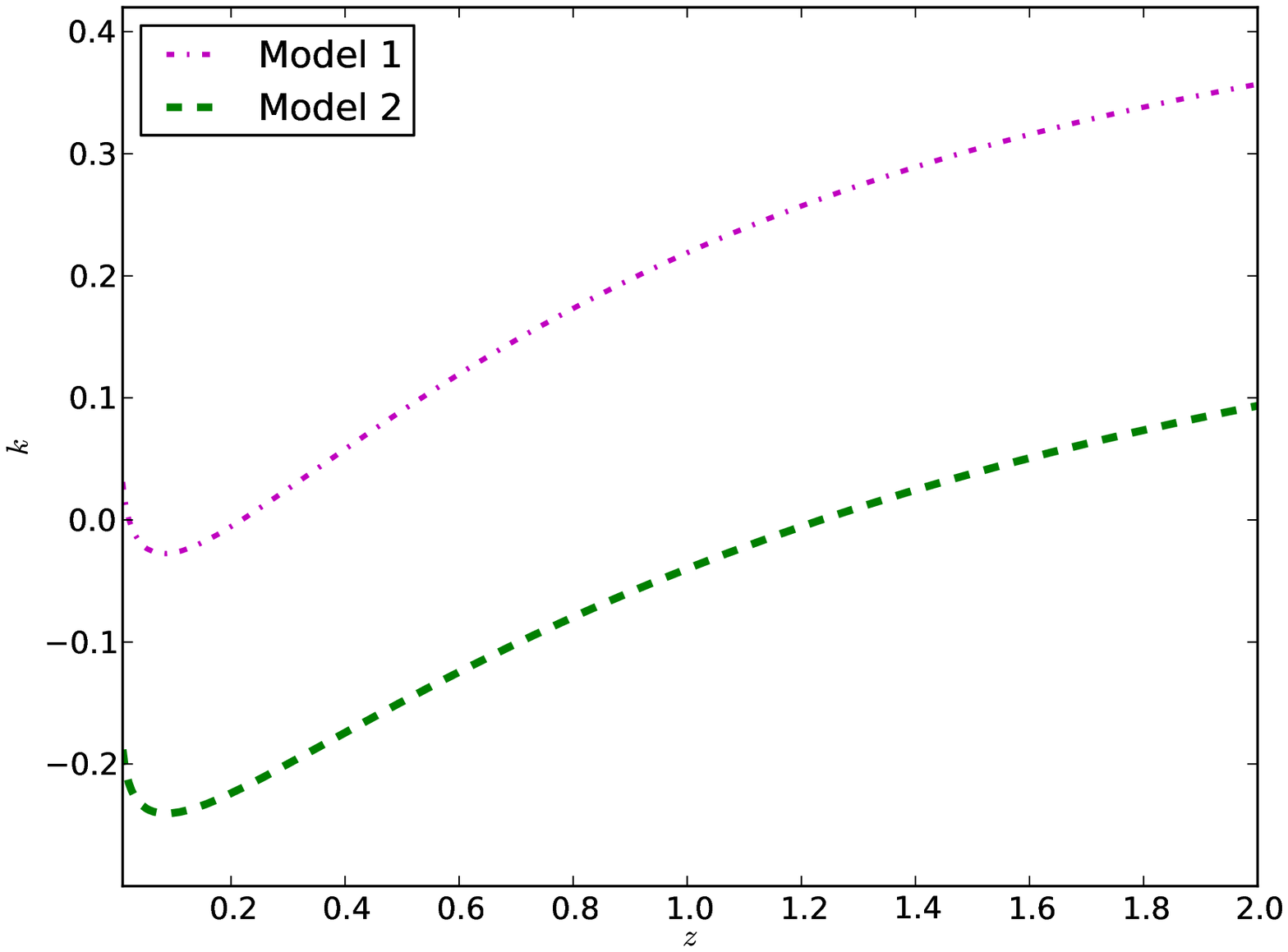}}
\begin{center} {\bf (a)} \end{center}
\end{minipage}
\begin{minipage}[t]{7.7cm}
\scalebox{1.0}{\includegraphics[angle=0, clip=true, trim=0.6cm 0.7cm 1.7cm 1.4cm, width=\textwidth]{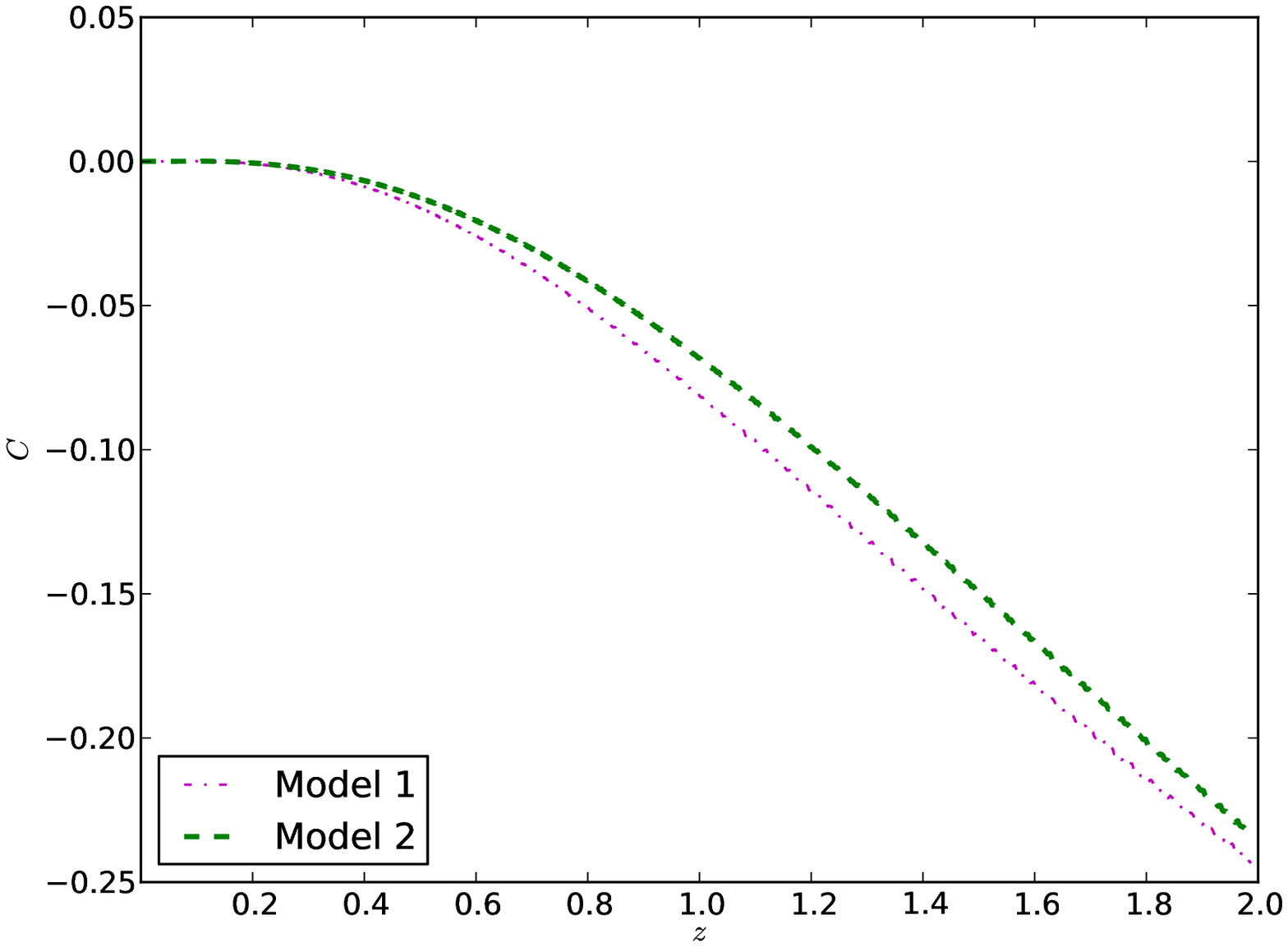}}
\begin{center} {\bf (b)} \end{center}
\end{minipage}
\caption{a) The FRW consistency parameter $k$. b) The FRW consistency parameter $\mC$.}
\label{fig:k}
\end{figure}

\para{Violation of the FRW consistency condition.}

The difference between the equations of state $\wHtot$
and $\wDtot$ is related to the redshift dependence of $k$,
shown in \fig{fig:k}a. For model 1, $k$ is positive
everywhere except for a small redshift interval close to
$z=0.2$, whereas for model 2, $k$ is positive at  large
$z$ and negative for $z<1.3$.
At $z=0$, $k=0.04$ for model 1 and $k=-0.20$ for model 2.
At $z\gg1$, $k$ approaches a constant as
backreaction becomes negligible, with asymptotic value $-0.2$
in model 1 and $0.2$ in model 2, and $|k|<0.5$  at all $z$.
A negative $k$ corresponds to $d'\geq 1/h$, so the distance
grows faster with increasing $z$ than in the spatially
flat FRW model with the same expansion history.
In well-fitting toy models, $k$ can have either sign, 
with a typical magnitude of $0.1\lesssim|k|\lesssim1$ at all $z$.
We expect the magnitude to be similar in the real universe
if the backreaction conjecture is correct.
However, a definitive test would require a reliable
calculation of the average expansion rate and more careful
\begin{wrapfigure}[16]{r}[0pt]{0.5\textwidth}
\scalebox{0.5}{\includegraphics[angle=0, clip=true, trim=0.7cm 0.7cm 1.7cm 1.3cm, width=1.0\textwidth]{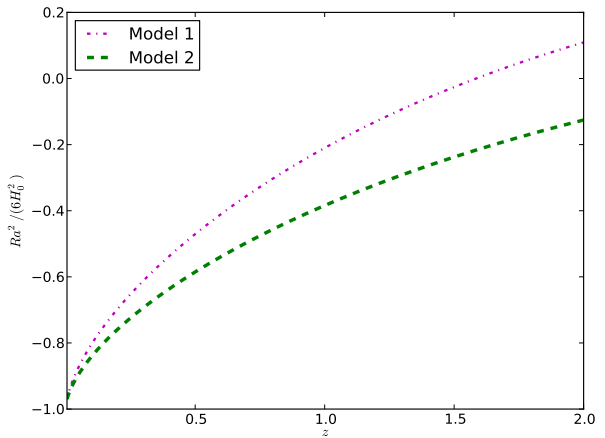}}
\caption{The quantity $\sR a^2/(6 H_0^2)$ that would be
constant and equal to $k$ in a FRW universe.}
\label{fig:curv}
\end{wrapfigure}
treatment of light propagation leading to equations
\re{z} and \re{DA} for redshift and angular diameter distance.
In \fig{fig:k}b we show $\mC$ that quantifies the redshift
dependence of $k$.

In the FRW case, $k=-\OKn$ is given by the spatial curvature,
but that is not the case if backreaction is significant.
In \fig{fig:curv} we show $\sR a^2/(6 H_0^2)=-a^2 h^2\OR$,
which in the FRW case would be equal to $k$.
For the backreaction models its evolution is quite
different from that of $k$, demonstrating that in general
$k$ does not describe spatial curvature. Even at large $z$
where backreaction is small and both $\sR a^2/(6 H_0^2)$
and $k$ are constant, their values (or even signs) do not in
general agree.


\subsection{Low redshift expansion} \label{sec:series}

\para{Distance bias.}

Let us consider the change of the distance--expansion rate
relation analytically. Unlike in the FRW case, there
is no closed form solution for the distance in terms of
the expansion rate, so we consider a series expansion around $z=0$.
Given an expansion history $h(z)$, the FRW distance--expansion rate
relation \re{dAFRW} gives
\bea \label{serFRW}
  d_A &\simeq& z - \frac{1}{2} ( 3 + q_0 ) z^2 + \frac{1}{6} \left( 11 + 7 q_0 + 3 q_0^2 - j_0 + \OKn \right) z^3 + \sO(z^4) \ ,
\eea

\noindent where $q_0=h_0'-1$ and $j_0=h_0''+(h_0'-1)^2$.
Solving for the distance from the backreaction relation
\re{dA} instead, we obtain
\bea \label{serBR}
  d_A &\simeq& z - \frac{1}{2} ( 3 + q_0 ) z^2 + \frac{1}{6} \left( 12 + 8 q_0 + 3 q_0^2 - j_0 - \frac{3}{2} \Omn \right) z^3 + \sO(z^4) \ .
\eea

Biasing of the distance shows up only at order $z^3$, like spatial curvature.
This can be easily seen from \re{dA}, as the only difference
between FRW and backreaction is the pressure term on the right-hand
side of the equation, and it doesn't contribute until at order $z^3$.
The difference at order $z^3$ corresponds to the jerk parameter
being shifted as $j_0\rightarrow j_0 - (1+q_0) + \frac{3}{2}\Omn + \OKn$.
If $\OKn=0$, we can write this as
$j_0\rightarrow j_0 - \frac{3}{2} (1-\Omn) (1+w_{H0})$.
So, assuming that the effective energy density of the extra component
is positive ($\Omn<1$), the distance for small $z$ is longer
in the backreaction case than in the FRW case if $w_H>-1$ and shorter
if $w_H<-1$. In other words, the distance is biased towards
the \LCDM model \cite{revs}. If the expansion history
is exactly that of the \LCDM model (plus possibly spatial
curvature), the distance is the same as in the FRW case.

It has been noted that because of the bias, the observed distances do
not necessarily imply that the expansion would have accelerated,
unlike in the FRW case \cite{Rasanen:2008b, Rasanen:2009b, revs, Buchert:2011}.
The expansion \re{serBR} shows that if
low redshift observations are analysed
with a power series expansion up to second order in $z$
and the coefficient of the second order term of $d_A$ is
found to be larger than $-\frac{3}{2}$, we can conclude that
the volume expansion rate has accelerated, just as in the FRW case.
(For observations extending to high redshifts, the
biasing cannot be estimated so simply, as we have seen.)
The small redshift expansion is limited by the fact that
the mean quantities $z$ and $d_A$ are not expected to describe
the physical redshift and distance over distances smaller than
the homogeneity scale, and an expansion of local quantities in
terms of the redshift is not useful due to large local variations.
This issue is present even if the average properties of the
universe are on large scales well described by the FRW model,
and deviations from the mean expansion on short distances
are usually known under the name peculiar velocities; see
\cite{Neben:2012} on limitations of the series expansion.
The homogeneity scale of 100 Mpc corresponds to $z=0.02$,
and data on such small redshifts is usually dropped due
to large local variations.
However, the homogeneity scale refers to a three-dimensional
average, and one-dimensional integrals like the redshift and the
distance may be more sensitive to inhomogeneity, depending on the
filamentarity of the matter distribution.

\para{FRW consistency parameter $k$.}

The series expansion also sheds some light on the
magnitude of the FRW consistency parameter $k$.
Using the expansion \re{serBR}, the FRW consistency parameter
\re{k} at $z=0$ can be written as
\bea \label{k0}
  k_0 &=& - q_0 - 1 + \frac{3}{2} \Omn = \Omn - 2 \OQn - 1 \ .
\eea

\noindent The amplitude of $k_0$ equals the deviation
from the \LCDM relation between $q_0$ and $\frac{3}{2}\Omn$.
With \re{hfit} and \re{k}, we can write $k=(1-h^2/\hfit^2)d^{-2}$,
so for $z\ll1$ we have $h/\hfit\simeq 1-\frac{1}{2} k_0 z^2$.
Observationally, we know model-independently that, roughly,
$0.2\lesssim\Omn\lesssim0.4$
\cite{Vonlanthen:2010, Audren:2012, Peebles:2004, Hubble, Riess:2011, Freedman:2012}.
Values of $q_0$ quoted in the literature depend strongly on the
assumed parametrisation \cite{moreacc, lessacc, trans, kinematic},
and are heavily based on distance measurements, whereas
it is the expansion rate which is of interest here.
From measurements of the expansion rate
\cite{Gaztanaga:2008, Blake:2011, Reid:2012, Blake:2012, Chuang:2012, Xu:2012, BAOrad, ages, Jimenez, Moresco:2012, Zhang:2012}
it is not yet possible to determine $q_0$ reliably,
but $|q_0|\lesssim1$ is a reasonable estimate.
Without a reliable calculation of backreaction, it is not
possible to definitely exclude the possibility that
$1+q_0$ and $-\frac{3}{2}\Omn$ accurately cancel.
However, as there does not appear to be any reason for such a
close cancellation, $k_0$ can be expected to be of the same order
as $1+q_0$ and $\Omn$ or slightly smaller, roughly
$0.1\lesssim|k_0|\lesssim1$.
In the toy model, this is the typical magnitude at all $z$, though
in model 1, $k_0$ is an order of magnitude smaller than $1+q_0$
and $\Omn$.

For FRW models, the value $k=-\OKn$ is tightly
constrained to be $|k|\lesssim0.01$ \cite{Okouma:2012}.
The strong constraint on $k$ is related to the fact that $k$ has
two roles in FRW models. It changes the expansion rate
by $h^2\rightarrow h^2-k(1+z)^2$ and affects the distance
by changing the spatial geometry, which is manifested
in the $\sinh$ behaviour in \re{dAFRW}.
(From the point of view of the light propagation equation \re{dA},
the latter effect follows from the former.) In \cite{Clarkson:2011d}
these effects were studied separately by introducing two parameters,
one that only modifies the spatial geometry and
another that only alters the expansion rate.
Taking the energy density to consist of dust and vacuum energy,
limits for the geometrical constant are $-0.4<k_0<1.2$ from the
Union2 data alone and $-0.1<k_0<0.2$ when other data are included.
(We quote error bars as 68\% C.L. limits and ranges as 95\% C.L. limits.)
Using the SDSS SN dataset instead of Union2, the limits
change somewhat, to $-0.8<k_0<1.2$ for SN data alone and
$-0.3<k_0<0.0$ with other datasets added.
The limits for the constant which modifies the expansion rate
are similar when only the SN data are used, and a factor
of 6 tighter with other data added.
If the two curvature parameters are taken to be the same, the
limit reduces to $|k_0|\lesssim0.01$ for the full dataset
(again assuming dust and vacuum energy), regardless of which set of
SNe is considered. With current CMB data, a similar
constraint can be achieved without any other datasets due to
the sensitivity of the late Integrated Sachs--Wolfe (ISW) effect
and lensing to the expansion rate at late times \cite{Ade:2013}.


In general, backreaction does not reduce to a modification
of the FRW spatial curvature constant, and if it does,
it follows from \re{dA} that the parameter appearing in
the $\sinh$ term and the expansion rate is the same,
unlike argued in \cite{Clarkson:2011d}.
If the average spatial curvature (or rather
$\av{\sR}+3\sQ$; see eq. 40 of \cite{Rasanen:2008b}) is
proportional to $a^{-2}=(1+z)^2$, the relation between
distance and expansion rate solved from \re{dA}
contains a $\sinh$ term. If the relative deviation
from the EdS behaviour drops down faster at large redshift,
the modification to the distance will be less extreme.
The tight constraint on spatial curvature in FRW models
(strongly driven by the measurement of the angular
diameter distance to last scattering from the CMB)
shows that the expansion rate has to approach the EdS case faster
than $(1+z)^2/\av{\rho}\propto (1+z)^{-1}$ as $z$ grows.
This constrains the value of $k(z)$ at high $z$, not at $z=0$.



To obtain a model-independent measure of $k$, distance and
expansion rate have to be observed independently.
The clustering of large-scale structure, in particular the radial
part of the BAO, contains detailed information about the expansion rate
\cite{Gaztanaga:2008, Blake:2011, Reid:2012, Blake:2012, Chuang:2012, Xu:2012, BAOrad},
and the expansion rate can be also determined from galaxy ages
\cite{ages, Jimenez, Moresco:2012, Zhang:2012}.
The resulting constraint is roughly $|k|\lesssim1$
\cite{Shafieloo:2009, Mortsell:2011}, much weaker than
when the FRW model is assumed, though the analyses
have not included all current data.
The constraint is within an order of magnitude of the expected
amplitude of $k$ from backreaction, so near future observations
may provide useful constraints.

There have been tests of the distance duality
relation $D_L=(1+z)^2 D_A$, where the angular diameter distance
data is obtained from BAO or galaxy ages \cite{duality}.
The FRW distance--expansion rate relation \re{dAFRW}
has been used to convert the expansion rate from galaxy ages and the
radial BAO component into distance. Apparent violation
of the duality relation could therefore be indicative of the failure
of the FRW distance--expansion rate relation instead.
No statistically significant discrepancy has been found.

\subsection{Other observations} \label{sec:otherobs}

\para{Expansion rate.}

\begin{wrapfigure}[19]{r}[0pt]{0.5\textwidth}
\scalebox{0.5}{\includegraphics[angle=0, clip=true, trim=1cm 0.7cm 1.7cm 1.3cm, width=\textwidth]{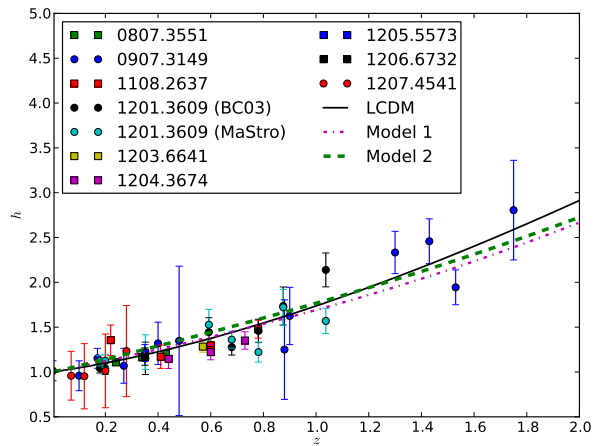}}
\caption{The Hubble parameter relative to its present value.
The data points are from large-scale structure (boxes)
\cite{Gaztanaga:2008, Blake:2011, Reid:2012, Blake:2012, Chuang:2012, Xu:2012}
and galaxy ages (circles)
\cite{Jimenez, Moresco:2012, Zhang:2012}.
Numbers refer to arXiv identifiers, labels BC03
and MaStro refer to stellar population synthesis models used
in \cite{Moresco:2012}.
}
\label{fig:h}
\end{wrapfigure}

We have fitted the toy model only to the SN
distance data. Let us now consider other observables.
In \fig{fig:h} we show the expansion rate $h(z)=H(z)/H_0$ for the two
backreaction models and the \LCDM model, along with data points
from large-scale structure
\cite{Gaztanaga:2008, Blake:2011, Reid:2012, Blake:2012, Chuang:2012, Xu:2012}
and galaxy ages \cite{Jimenez, Moresco:2012, Zhang:2012}.
We have adopted the value $H_0=72$ km/s/Mpc, and the points
could be moved up or down together by the uncertainty in $H_0$.\footnote{The
value $H_0=73.8\pm2.4$ km/s/Mpc is quoted in \cite{Riess:2011},
and the more recent study \cite{Freedman:2012} gives $H_0=74.3\pm2.1$.
We allow for the possibility of unaccounted systematics and simply take
the upper and lower ranges obtained with the three different
calibrators used in \cite{Riess:2011}, 64 km/s/Mpc $<H_0<$ 81 km/s/Mpc,
or $H_0=72\pm4$ km/s/Mpc. The precise central value and error bars have
little effect on our results.}
The assigned errors of some galaxy age measurements have been
criticised as too small \cite{Mortsell:2011}, there is some
model-dependence in the large-scale structure analysis
and the large-scale structure data points are not all independent.
Therefore, the data points and their errors in \fig{fig:h} are only
indicative. The goodness-of-fit is $\chi^2=17$ for the \LCDM model,
$\chi^2=18$ for model 1 and $\chi^2=19$ for model 2, for 37 data points
(for either choice of stellar population synthesis model used in
\cite{Moresco:2012}).
This agrees with the visual impression that differences between
the \LCDM model and the backreaction models are not large compared
to the error bars.

\para{Age of the universe.}

We can also consider the age of the universe relative
to the Hubble parameter.
Globular clusters give the 95\% lower limit $t_0>$ 11.2 Gyr and
the best-fit age 13.4 Gyr \cite{Krauss:2003}. Taking
$t_0=13.4\pm1.1$ Gyr and $H_0=72\pm4$ km/s/Mpc would give
$H_0t_0=0.99\pm0.14$. However, the distribution of $t_0$ is not
Gaussian, and the ``firm upper limit'' is given as $t_0<$ 21 Gyr.
Considering only the lower limit $t_0\geq$ 11.2 Gyr and fixing 
$H_0=72$ km/s/Mpc gives $H_0t_0\geq0.82$.
There is a more recent and higher age estimate of $14.61\pm0.8$ Gyr,
which is however based only on a single star \cite{Verde:2013}.
With $H_0=72$ km/s/Mpc, this would give $H_0t_0=1.08\pm0.06$.
Figure \ref{fig:Ht} shows that the backreaction models
satisfy the constraint $H_0t_0\geq0.82$ easily: $Ht$ is close to
unity today because the volume is almost completely taken up by
underdense region, which is almost completely empty.
In the real universe, the value of $H_0t_0$  is not expected
to be as close to unity as in the toy model.
In backreaction models, $Ht<1$ is a general prediction, provided
matter can be treated as dust and vorticity is not important
\cite{Rasanen:2005, Rasanen:2009b}.
Were the value $H_0t_0=1.08$ to be reliably confirmed with more
precision, it would rule out the backreaction conjecture.
(In combination with other cosmological datasets, it would also rule
out the \LCDM model.)

\para{Distance to the last scattering surface.}

The main sensitivity of the CMB to late-time evolution of the
universe is via the angular diameter distance, which affects the
position of the acoustic peaks in multipole space \cite{Vonlanthen:2010}.
The Union2.1 data gives distances relative to $H_0$,
whereas the CMB peak position gives the distance to
the redshift of the last scattering surface at
$z_*\approx1090$ in physical units\footnote{More precisely, the peak
position fixes $D_A(z_*)$ relative to the sound horizon at last
scattering, but the sound horizon is also determined by the CMB data.}.
Therefore, any model fitted to only the Union2.1 data is consistent
with either any value of $D_A(z_*)$ or any value of $H_0$,
but the combination of all three measurements imposes constraints.
Model-independent analysis of WMAP seven-year data and SPT data
gives $D_A(z_*)=12.7\pm0.2$ Mpc \cite{Vonlanthen:2010, Audren:2012}.
(The precision of current observations of $D_A(z_*)$ relative to the
sound horizon from the Planck satellite is a factor of 1.7 better,
but they are also more model-dependent \cite{Ade:2013}.)
With $H_0=72\pm4$ km/s/Mpc, this translates into $d(z_*)=3.3^{+0.3}_{-0.2}$.
We have simply combined the 68\% ranges, so the error is rather conservative.
For the \LCDM best-fit model we have $d(z_*)=3.3$, whereas models
1 and 2 give $d(z_*)=0.72$ and $d(z_*)=1.6$, respectively,
in strong disagreement with observation.

The fact that the \LCDM model fitted to distance data at small
redshift correctly predicts the distance to $z\approx1090$
is indicative of the fact that any model that fits the low $z$
observations and rapidly approaches EdS behaviour at $z\gtrsim2$ (with a
radiation contribution at high $z$) will fit all data also at higher $z$.
This is not the case for our toy models.
The expansion rate at early times is smaller than in the EdS case,
so there is less overall expansion, leading to shorter distances.
This is related to the fact that as the toy model is constructed
as the union of two FRW models, the contribution of spatial curvature
relative to matter density decreases slowly with increasing redshift.
Figure \ref{fig:Ht}b shows that the expansion rate in the toy model deviates
noticeably from the EdS behaviour already at $10^{-2} t_0$, corresponding
to less than a billion years of age, a time when the \LCDM expansion rate is
indistinguishable from the EdS case.
The problem could be fixed by tweaking the expansion history
to make the transition from EdS behaviour sharper, but
that would be taking the toy model too seriously.
If backreaction is significant in the real universe, the timescale
of the change from EdS behaviour is related to structure formation.
In a semirealistic model based on statistics of structure formation
the timescale is not dissimilar to that shown in \fig{fig:Ht}
\cite{peak} (see figure 1 in \cite{Buchert:2011}).
What happens in the fully realistic case remains to be determined.

In addition to the peak positions, with current CMB data the lensing
and the ISW effect (in the backreaction context more properly called
the Rees--Sciama effect due to its non-linear origin)
have become important cosmological probes, as they are
affected by the expansion rate at late times \cite{Ade:2013, Das:2013}. 
Even if backreaction is significant, such secondary anisotropies
are expected to be small in absolute terms despite large local
variations in the geometry, because of statistical homogeneity and isotropy
\cite{Rasanen:2008b, Rasanen:2009b, Rasanen:2009a}, but their magnitude
relative to the primary anisotropy remains to be determined.
The ISW contribution to the overall two-point correlation function
does not significantly deviate from the \LCDM prediction, but
there is excess power along lines of sight that cross extrema of the
density field \cite{ISWpower}, and there are also other anomalies
on large angular scales \cite{asym}.

\section{Conclusion} \label{sec:conc}

\para{Summary.}

Backreaction, i.e. change of the average expansion rate due to structure formation, is a possible explanation for the observation that the
expansion rate and distance are larger than predicted by
homogeneous and isotropic models with ordinary
matter and ordinary gravity. So far there is no fully
realistic calculation of the effect of structure formation
on the average expansion rate of the universe.
Nevertheless, it is possible to test backreaction by comparing
measurements of distance and expansion rate, as their relation
is different than in the FRW case \cite{Rasanen:2008b, Rasanen:2009b}.

We have presented a toy model of backreaction where the
expansion rate has realistic features. We have shown
that the model provides a good fit to the Union2.1 SN
data, even though the expansion history is quite different
from the \LCDM model. The model also illustrates the feature
that the effective equation of state inferred from distance
measurements is different from the effective equation of state
that corresponds to the expansion rate.
We find that the deviation from the FRW distance--expansion rate relation
in the toy model is slightly below current observational constraints.
The order of magnitude is expected to be the same in the real
universe if backreaction is signiﬁcant. This provides a distinct signature
that may be used to put meaningful constraints on backreaction with
improved observations of the expansion rate in the near future.
We have also defined a distance--expansion rate consistency condition
for backreaction that may be used to test backreaction.
However, improved theoretical treatment is needed for
definite tests, and the average expansion rate, light
propagation and structure formation all need to be understood better.

\acknowledgments

SR thanks Bruno Leibundgut for correspondence, and we thank Phil Bull
for pointing out a binning error in a previous version of figure 1.

\end{document}